# QoE Management of Multimedia Streaming Services in Future Networks: A Tutorial and Survey

Alcardo Alex Barakabitze, Nabajeet Barman, Arslan Ahmad, Saman Zadtootaghaj, Lingfen Sun, Maria G. Martini, Luigi Atzori

*Abstract*—The highly demanding Over-The-Top (OTT) multimedia applications pose increased challenges to Internet Service Providers (ISPs) for assuring a reasonable Quality of Experience (QoE) to their customers due to lack of flexibility, agility and scalability in traditional networks. The future networks are shifting towards the cloudification of the network resources via Software Defined Networks (SDN) and Network Function Virtualization (NFV). This will equip ISPs with cutting-edge technologies to provide service customization during service delivery and offer QoE which meets customers' needs via intelligent QoE control and management approaches. Towards this end, we provide in this paper a tutorial and a comprehensive survey of QoE management solutions in current and future networks. We start with a high-level description of QoE management for multimedia services, which integrates QoE modelling, monitoring, and optimization. This followed by a discussion of HTTP Adaptive Streaming (HAS) solutions as the dominant technique for streaming videos over the best-effort Internet. We then summarize the key elements in SDN/NFV along with an overview of ongoing research projects, standardization activities and use cases related to SDN, NFV, and other emerging applications. We provide a survey of the state-of-the-art of QoE management techniques categorized into three different groups: a) QoE-aware/driven strategies using SDN and/or NFV; b) QoE-aware/driven approaches for adaptive streaming over emerging architectures such as multi-access edge computing, cloud/fog computing, and information-centric networking; and c) extended QoE management approaches in new domains such as immersive augmented and virtual reality, mulsemedia and video gaming applications. Based on the review, we present a list of identified future QoE management challenges regarding emerging multimedia applications, network management and orchestration, network slicing and collaborative service management in softwarized networks. Finally, we provide a discussion on future research directions with a focus on emerging research areas in QoE management, such as QoE-oriented business models, QoE-based big data strategies, and scalability issues in QoE optimization.

## I. INTRODUCTION

Multimedia services consumption has increased tremendously over the past years and is expected to continue to grow even more over the next years. According to the latest Cisco Visual Networking Index (VNI) Forecast [1], 82% of all IP traffic will be video by 2022. Mobile-connected devices (e.g., Device to Device (D2D) and Machine to Machine (M2M) communications) are estimated to be 14.6 billion by 2022 and therefore exceeding the world's projected population of 8 billion by 2022. This exponential growth, due to the increasing popularity and use of video streaming services (e.g., YouTube and Netflix), has triggered and introduced new revenue potential for Internet Service Providers (ISPs), mobile operators and Over-The-Top providers (OTTP). Delivering high video quality to the end users is very important for the continued success of such services [2], [3]. Today, the end-users are accustomed to more resource demanding services with better quality from ISPs [4], [5]. However, achieving good Quality of Experience (QoE) is a challenging task because of many factors such as different client devices/request patterns, changing media contents, varying transmission/network conditions, and significant spatial and temporal variation in the performance of Content Distribution Networks (CDNs). Great efforts from both academia and industry have been made to optimize the video content delivery chain and enhance end users' QoE. The most common mechanisms used for improving end-users' QoE are either based on network optimization (e.g. QoE-driven network resource allocation and QoE-driven routing) or client-driven adaptive video streaming [6]–[9].

Despite these efforts, QoE management remains a challenging task due to many issues [8], [10] which can be categorized into four different aspects, as illustrated in Fig. 1. The first aspect is the variability of network resources, unstable nature of wireless channels and the characteristics of fixed/mobile networks in heterogeneous environments [11]. Congested areas such as trains, stadiums and shopping malls require continuous adaptation of network resource allocation to various clients. The second aspect is the emergence of new services (e.g., video gaming and virtual/augmented reality (VR/AR)), the diversity of context of use, the users' expectations combined with the operational cost optimization by mobile and service providers. The third aspect is the consideration of a variety of networks (e.g., fixed and mobile), where different measurement and assessment methods are required to be employed for QoE management considering resource constraints. The final aspect is associated with the popularity and fast growth of the multimedia services over the Internet, and the heterogeneity of end-user devices with different capabilities (e.g., screen size, computational power/resources, and storage capabilities). This poses even more challenges when allocating resources among users with different QoE preferences. Fig. 1 summarizes

Alcardo Alex Barakabitze and Lingfen Sun are with Plymouth University, U.K., Email: {alcardoalex.barakabitze, l.sun}@plymouth.ac.uk
Nabajeet Barman and Maria Martini are with Kingston University, U.K. Email: {n.barman, m.martini}@kingston.ac.uk
Arslan Ahmad is with IS-Wireless, 05-500, Piaseczno, Poland. Email: a.ahmad@is-wireless.com
Saman Zadtootaghaj is with Technische Universität Berlin, Germany. Email: saman.zadtootaghaj@qu.tu-berlin.de
Luigi Atzori is with University of Cagliari, Italy. Email: l.atzori@ieee.org

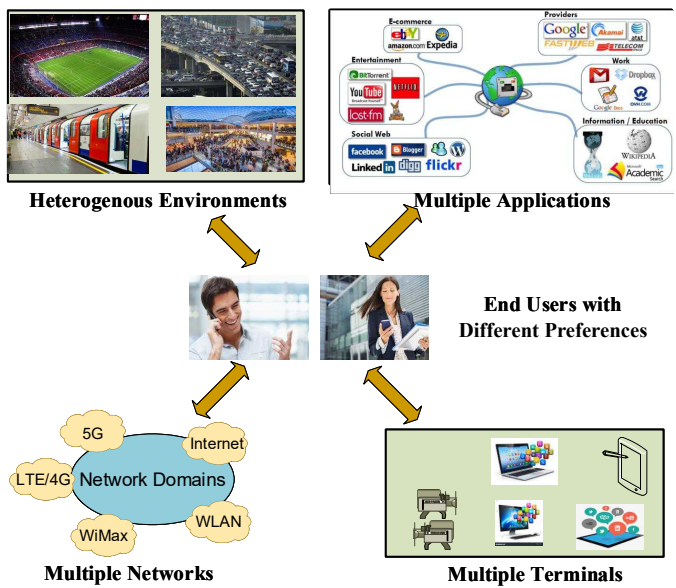

Fig. 1: QoE management challenges in future networks.

the QoE management challenges in IP-based networks. The increased user demand and expectations of services with excellent quality have triggered telecom operators to upgrade their systems and invest in new cutting-edge network softwarization paradigms such as Software Defined Networking (SDN) [12], Network Function Virtualization (NFV) [13], Multi-Access Edge Computing (MEC) [14] and Cloud/Fog Computing (C/FoC). This transformation and the upgrade of their systems is also driven by the mounting pressure of new emerging use cases, ranging from ultra-high definition video resolutions (4K/8K), network-controlled D2D communications, Machine Type Communication (MTC) and Massive Internet of Things (MIoT).

In such an agile and flexible environment, it is essential to consider new solutions, such as the separation of user and control planes, and possibly re-define the boundaries between the network domains (e.g., radio access network and the core network). Therefore, there is a need for new advanced autonomic network management platforms that can guarantee the end users' QoE, especially in heterogeneous environments. Towards this end, new paradigms such as SDN and NFV have been identified as critical technologies for enabling future network control to be programmable, centrally manageable, adaptable and cost-effective. SDN and NFV are considered to be suitable for applications such as video streaming [15]–[22]. Indeed, SDN and NFV may enable network management to be automated and ensure that the end users' QoS/QoE requirements as well as Experience Level Agreement (ELA)[1] are fulfilled in the heterogeneous environments [23].

More importantly, SDN and NFV can provide end-to-end resource, infrastructure and services across multi-programmable domains that belong to different operators or service providers. To this end, investigations are ongoing as to what extent future networks and different supporting technologies can be software-configurable and software platforms to be hardware-agnostic. Aligned with the continuing SDN and NFV research activities by some standardization bodies, there is an urgent need to study and explore the adoption of SDN and NFV on how they could introduce significant changes on the operation, management, and delivery of QoE-aware services in the context of emerging and future networks.

*A. Related Work and Motivation*

The industry and the academia are embracing SDN and NFV at an unprecedented speed as future potential technologies to provide service customization and better solutions concerning QoE control and management of multimedia services in current and future networks (e.g., 5G) [19], [24]. To this end, different works in the past have been proposed to identify potential approaches, use cases and architectures for QoE management of multimedia services in fixed and mobile networks. Baraković and Skorin-Kapov [8] presented a survey of QoE management in wireless networks by focusing on three broad management aspects, namely, QoE modeling, QoE measurement, and QoE adaptation. Skorin-Kapov *et al.* in [25] discussed emerging concepts and challenges related to managing QoE for networked multimedia services. By focusing on parameters that are taken into account for QoE optimization, surveys on QoE-based scheduling strategies for wireless systems were presented by Sousa *et al.* [26] and by authors in [27], [28] and [29]. Seufert *et al.* [6] presented a survey on QoE in HTTP adaptive video streaming, while Barman and Martini [30] provided a comprehensive survey of various QoE models for HTTP Adaptive Streaming (HAS) applications. Liotou *et al.* [10] provided insights regarding network-level QoE management in mobile cellular networks. Petrangeli *et al.* [31] have recently surveyed the QoE-centric management of adaptive video streaming services. A survey on existing work on QoE modeling and assessment (including both subjective and objective) and QoE management of video transmission over various types of networks is provided by Zhao *et al.* [32] and Su *et al.* [33]. However, the survey papers mentioned above do not explain how QoE can be improved or optimized using SDN and NFV. As such, these studies do not provide insights into how network or service providers can utilize the capabilities of SDN and NFV for QoE management (e.g., to achieve automation, programmability, flexibility, scalability and network control) to meet the end users' quality demands.

Different surveys attempt to fill in the mentioned gaps by identifying various approaches, uses cases, architectures and the huge benefits brought by SDN and/or NFV for QoE management of multimedia services in the perspective of current and future networks [19], [24], [34]. Wang *et al.* [19] presented a review on the state-of-the-art of QoE from several perspectives (e.g., assessment methods, QoE models, and control methods) in SDN/NFV-based networks. QoE-driven and energy-aware video adaptation use cases in future SDN/NFV-based networks, as envisaged by the SELFNET project, are presented in [24]. Peng *et al.* [35] gave a discussion on QoE-oriented mobile edge service management using SDN and NFV, while Awobuluyi *et al.* [21] provided a discussion on the context-aware QoE management in the SDN control perspective of future networks. The works in [18] and [34] presented different approaches for the autonomic management in an SDN/NFV networking paradigm. Table I provides a summary of related survey papers on the aspect of QoE management of multimedia streaming services.

---

[1] Experience Level Agreement (ELA): Indicates a QoE-enabled counterpart to traditional QoS-based Service Level Agreement (SLA) that conveys the performance of the service in terms of QoE. The ELA establishes a common understanding of an end-user's experience while using the service.

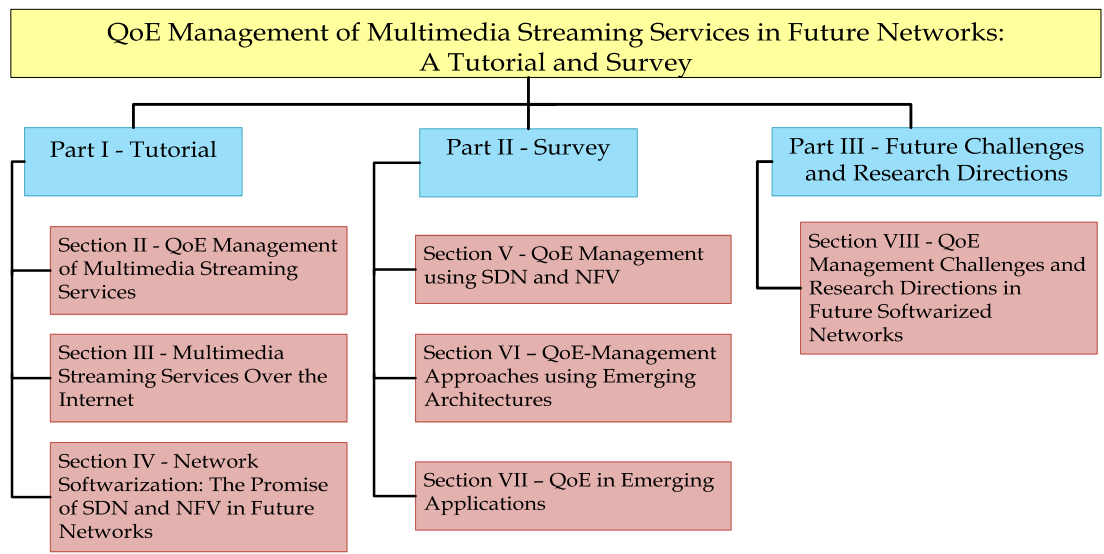

Fig. 2: Structure and organization of the paper.

Although recent efforts explore the QoE management aspects using SDN and/or NFV, we note that these works are limited in at least one of the following: 1) limited review and discussion of the standardization activities related to SDN and/or NFV; 2) lack of comprehensive descriptions of ongoing research projects, state-of-the-art efforts, challenges as well as concrete research directions in SDN and/or NFV, and emerging multimedia services and applications in future networks; 3) with regard to scope, they do not provide different architectural approaches, implementations and deployment strategies for QoE management using SDN and/or NFV; 4) they do not describe how QoE can be managed/measured or controlled in emerging technologies (such as cloud/fog computing, MEC or Information Centric Networking (ICN)) and new domains with new constraints (e.g., delay sensitive applications such as AR/VR or video gaming applications).

### B. Scope and Contributions

The primary objective of this paper is to provide the reader with a comprehensive state-of-the-art regarding the QoE management of multimedia streaming services in future networks. The contribution of this work is composed of three parts:

1) *Tutorial* – We first provide a discussion on the critical components of QoE management: QoE modeling; QoE monitoring and measurement, QoE optimization and control for multimedia OTT streaming services. In light of the above, we also provide a discussion on HAS solutions as the dominant technique for streaming videos over the Internet. We also present a discussion on the server and network-assisted approach and critical issues in the multimedia service management. A tutorial of network softwarization and virtualization in the future Internet with the perspectives of multimedia streaming is also provided.

2) *Survey* – We present a survey of the state-of-the-art of QoE management techniques that leverage SDN and NFV paradigms. Some of the QoE management topics covered include the usage of the server and network-assisted architectures, OTTP and ISP collaboration, new applications and transport layer protocols in SDN/NFV networks. We further present a detailed description of QoE-aware/driven approaches for adaptive streaming over emerging architectures such as MEC, cloud/fog computing, and ICN. Moreover, we also extend QoE management approaches to new domains including immersive augmented and virtual reality, mulsemedia, light field and video gaming applications.

3) *Future Research Challenges and Directions* – We identify future challenges and research directions/recommendations concerning the QoE management of multimedia services in the context of future softwarized networks. We explore the QoE management challenges regarding emerging multimedia applications, network management and orchestration, network slicing and collaborative service management of the multimedia services in softwarized networks. The future research directions provide foresight on the QoE management in emerging research areas such as QoE business models, QoE-based big data strategies, security, privacy and QoE-based trust models and scalability issues for QoE optimization in future softwarized networks.

TABLE I: A Summary of related survey papers and systematic discussion.

| Survey Paper | Year | Topics Covered and Scope | SDN or/and NFV Considerations | QoE in Emerging Architectures | QoE in New Domains |
|---|---|---|---|---|---|
| Baraković et al. [8] | 2013 | QoE modeling, monitoring and measurement | No | No | No |
| Liotou et al. [10], Seufert et al. [6] | 2015 | QoE in HTTP adaptive video streaming [6], network-level QoE management in mobile networks [10] | No | No | No |
| Awobuluyi et al. [21] | 2015 | Context-aware QoE management in the SDN | SDN only | No | No |
| Zhao et al. [32], Su et al. [33] | 2016 | QoE assessment and management in video transmission [32], QoE of video streaming [33] | No | No | No |
| Wang et al. [19] | 2016 | Architecture for personalized QoE management | Yes | No | No |
| Peng et al. [35] | 2017 | QoE-oriented mobile edge service management | Yes | MEC only | No |
| Sousa et al. [26] | 2017 | QoE-based scheduling strategies | No | No | No |
| Skorin-Kapov et al. [25], Petrangeli et al. [31] | 2018 | QoE modeling, QoE monitoring and management [25], QoE-centric management of adaptive video streaming services [31] | To some extent | ICN only [31], MEC only [25] | AR/VR & multisensory |
| Barman and Martini [30] | 2019 | QoE modelling for HTTP adaptive video streaming | No | No | No |
| Barakabitze et al. [36] | 2019 | Network slicing using SDN and NFV | Yes | Yes | No |
| Our work | 2019 | (a) A tutorial on QoE modelling and assessment, QoE monitoring and measurement, QoE optimization and control; (b) a survey on QoE management in SDN and NFV, and (c) QoE management using emerging architectures and in new domains | Yes (both SDN and NFV) | Yes (MEC, fog/cloud computing and ICN) | Yes |

\*\* QoE in New Domains: We refer to aspects of QoE in new domains such as AR/VR, mulsemedia and gaming video streaming applications. It is to be noted that with the exception of Skorin-Kapov *et al.* [25] and Petrangeli *et al.* [31] which discuss AR/VR & multisensory applications, other survey papers are limited in this aspect which is covered in this work.\*\*

## C. Paper Structure and Organization

The paper is organized as follows: The *tutorial* discussion on QoE management is presented in Section II along with the definition of QoE and critical components of QoE management. Section III provides a tutorial on multimedia streaming solutions over the Internet including the discussion on HAS solutions, server and network assisted DASH, and key issues in multimedia service delivery and management. Section IV discusses the current network softwarization (SDN and NFV) strategies and relevant advancements in the next generation networks for multimedia services delivery. Section V provides the *survey* of the state-of-the-art works on QoE-centric management in the future Internet using SDN and NFV. Section VI presents a survey of QoE management using emerging technologies such as mobile edge computing and cloud computing. Section VII provides a comprehensive description of QoE in emerging applications such as immersive AR/VR, cloud gaming, light field display and mulsemedia. Section VIII investigates the *future research challenges and directions* related to QoE management. We finally provide our concluding remarks in Section IX. For a better understanding of the structure and organization of this paper, we refer the readers to Fig. 2. Table II shows a summary of the used acronyms in the paper.

## II. QoE Management for Multimedia Streaming Services

This section provides a tutorial on the aspects of QoE, QoE modeling and assessment along with some metrics and important KPIs used for QoE management evaluation. It also provides an overview of QoE monitoring and measurement, QoE optimization and control using the end-to-end multimedia services delivery chain. Fig. 3 represents the critical components of QoE management of multimedia services.

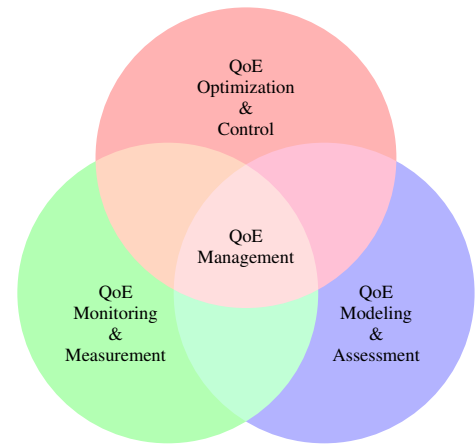

Fig. 3: Key components of QoE management for multimedia streaming services.

### A. Quality of Experience: Definition

In the past, QoS based measurements that consider network parameters (e.g., packet loss, delay and throughput) were used to define the level of satisfaction/performance of a service. As QoS metrics are not explicitly and directly linked with a customer's satisfaction of a service, user-centric Quality of Experience (QoE) metrics (e.g. Mean Opinion Scores) have been used in recent years to assess the quality of multimedia services, in addition to QoS metrics. QoE considers the user's subjectivity towards a specific service which can be defined as *"the degree of delight or annoyance of the user of an application or service. It results from the fulfillment of his or her expectations with respect to the utility and/or enjoyment of the application or service in the light of the users' personality and current state"* [37]. The understanding of the users' expectations and experiences from a service is vital for the success of a service.

TABLE II: LIST OF COMMONLY USED ACRONYMS IN THIS PAPER.

| Abbr. | Definition | Abbr. | Definition | Abbr. | Definition |
|---|---|---|---|---|---|
| AP | Action Plane | IoT | Internet of Things | PED | Parameters Enhancing Delivery |
| AR | Augmented Reality | IRTF | Internet Research Task Force | PER | Parameters Enhancing Reception |
| AS | Autonomous System | ISPs | Internet Service Providers | POMDP | Partially Observable Markov Decision Process |
| ATIS | Alliance for Telecommunications Industry Solutions | KPIs | Key Performance Indicators | PSNR | Peak Signal to Noise Ratio |
| BMS | Bandwidth Management Solution | KQIs | Key Quality Indicators | PBN | Policy/Intent Based Networking |
| CAPEX | CApital EXpenditure | M2M | M2M Machine to Machine | QFF | QoE Fairness Framework |
| CC | Cloud Computing | ML | Machine Learning | QoBiz | Quality of Business |
| CDNs | Content Distribution Networks | MTC | Machine Type Communications | QoE | Quality of Experience |
| cDVD | client-Driven Video Delivery | MANO | Management and Orchestration | QoS | Quality of Service |
| CMAF | Common Media Application Format | MIoT | Massive Internet of Things | QUIC | Quick UDP Internet Connections |
| CPP | Controller Placement Problem | MOS | Mean Opinion Score | RAN | Radio Access Network |
| D2D | Device to Device | MPD | Media Presentation Description | RR | Reduced Reference |
| DANE | DASH-Aware Network Elements | MEF | Metro Ethernet Forum | SQAPE | Scalable QoE-aware Path Selection |
| DASCache | Dynamic Adaptive Streaming over popularity-driven Caching | MEC | Multi-Access Edge Computing | SVC | Scalable Video Coding |
| DASH | Dynamic Adaptive Streaming over HTTP | MNO | Mobile Network Operator | SABR | SDN-assisted Adaptive Bitrate Streaming |
| DHCP | Dynamic Host Configuration Protocol | MPTCP | Multipath TCP | SIDs | Segment IDentifiers |
| DNS | Domain Name System | MULSEMEDIA | MULtiple SEnsorial MEDIA | SR | Segment Routing |
| DPI | Deep Packet Inspection | MPLS | Multi-Protocol Label Switching | SAND | Server and Network Assisted DASH |
| ELA | Experience Level Agreement | MSBN | Multi-Service Broadband Network | SFC | Service Function Chaining |
| ELBA | Efficient Layer Based Routing Algorithm | NAT | Network Address Translation | SLAs | Service Level Agreements |
| XML | Extensible Markup Language | NE | Network Element | SDN | Software Defined Networks |
| ETHLE | Edge-based Transient Holding of Live Segment | NFV | Network Function Virtualization | SBI | Southbound Interfaces |
| ETM | Economic Traffic Management | NMS | Network Management System | SSIM | Structural Similarity Index Metric |
| ETSI | European Telecommunication Standard Institute | NSC | Network Service Chain | SS | Surrogate Server |
| FD | Full-Duplex | NUM | Network Utility Maximization | TCAM | Ternary Content Addressable Memory |
| FoC | Fog Computing | NN | Neural Network | URL | Uniform Resource Locator |
| FR | Full Reference | NGN | Next Generation Networks | VMAF | Video Multi-Method Assessment Fusion |
| HAS | HTTP Adaptive Streaming | NR | No Reference | VMs | Virtual Machines |
| HDR | High Dynamic Range | ONF | Open Network Foundation | VNFs | Virtual Network Functions |
| HFC | Hybrid Fiber Coax | OSM | Open Source MANO | VPN | Virtual Private Networks |
| ICN | Information Centric Networking | OVF | Open Virtualization Format | VQO | Viewer QoE Optimizer |
| IFs | Influence Factors | OPEX | OPerational EXpenditure | VSSs | Virtual Surrogate Servers |
| ITU | International Telecommunication Union | OTT | Over-The-Top | VR | Virtual/Augmented Reality |
| IETF | Internet Engineering Task Force | OTTP | Over-The-Top Provider | WCG | Wide Color Gamut |
| IXPs | Internet eXchange Points | PGW | Packet Data Network Gateway | WAN | Wireless Access Network |

### B. QoE Modeling and Assessment: Metrics and Models

Based on the ITU-T Rec. P.10/G.100 Amendment 5 [37],QoE assessment can be defined as the process of measuring or estimating the QoE for a set of users of an application or service with a dedicated procedure and considering the influencing factors (possibly controlled, measured, or simply collected and reported). The QoE assessment, for example, based on SDN/NFV techniques [38]–[40] is one of the essential steps toward QoE-based monitoring and management. Depending on the objective and focus of the study, different subjective assessment methods and guidelines are described in the ITU-T Rec. BT.500 [41], P.910 [42] and P.913 [43].

Over the past years, the image and video research community have dedicated much efforts towards the development of metrics and models which can estimate the quality of an image or video objectively [44]–[46]. Such metrics use the image and video properties and try to predict the quality as would be perceived by the human. Depending on the amount of source information required, they can be classified as Full Reference (FR) (full source information required), Reduced Reference (RR) (partial source information required) and No-reference (NR) (no source information is required). Due to full access to source information, FR metrics are usually more accurate than RR and NR metrics. A review and classification of existing models and metrics proposed for QoE estimation for HTTP Adaptive Streaming (HAS) based applications is provided by Barman and Martini in [30] while a survey on QoE metrics and assessment methodologies is provided in [47]. For an overview of the QoE measurement approaches, we refer the reader to the survey and tutorial paper by Juluri *et al.* in [48]. Table III presents some of the most commonly and widely used quality metrics of image and video quality assessment.

### C. QoE Monitoring and Measurement

The process of managing and optimizing the end users' QoE requires knowledge regarding the root cause of QoE degradation or unsatisfactory QoE levels. In that respect, relevant information and data related to terminal capabilities (e.g., screen size, display performance), application/service specific information and its quantification, QoE-related information inside the network have to be monitored, collected and measured [59]. The network parameters, as well as bitstream information, can be collected from client devices or network elements using monitoring probes. With the advancement of virtualization and data management capabilities, virtualized probes [60] can be used to provide efficient QoE monitoring mechanisms in future softwarized networks. The collected network-level Key Performance Indicators (KPIs), e.g., throughput, packet loss, delay or user-level service/application specific Key Quality Indicators (KQIs), e.g., frame rate, video resolution, service usability, and reliability, provide inputs for QoE estimation models [48], [61].

### D. QoE Optimization and Control

The QoE management of multimedia services, as shown in Fig. 4, involves continuous optimization and dynamic control of relevant mechanisms, from content generation to content consumption, along the service delivery chain. One of the ultimate goals of QoE management is the maximization of the end users' QoE level through the efficient allocation of available network resources. However, considering the multimedia delivery chain shown in Fig. 4, QoE optimization and control is a challenging task due to many issues, including the heterogeneity of multimedia-capable users' devices. As stated in [8], the main challenges that arise with regards to QoE optimization and control may be summarized in the answers to the following four questions: (1) What key quality parameters to optimize and control? (2) Where to control (e.g., at the client, server and the network side)? (3) When to perform QoE optimization and control (e.g., during the service, that is, on-line control or in an off-line fashion [19])? (4) How often to control and optimize QoE? Different studies have been conducted in the literature to answer the above questions.

TABLE III: Some of the commonly used image and video QA models.

| Metric | Year | Model Type | Modality |
|---|---|---|---|
| Peak Signal-to-Noise Ratio (PSNR) [49] |  | FR | Images - Frames |
| Structural Similarity Index Metric (SSIM) [44] | 2004 | FR | Images - Frames |
| Video Multimethod Assessment Fusion (VMAF) [45] | 2016 | FR | Images - Frames |
| Visual Information Fidelity (VIF) [50] | 2006 | FR | Images - Frames |
| HDR-VDP-2: A calibrated visual metric for visibility and quality predictions in all luminance conditions [51] | 2011 | FR | HDR images |
| Video Quality Metric (VQM) [52] | 2004 | FR | Video |
| Reduced Reference Entropic Differencing (RRED) [53] | 2013 | RR | Video |
| Spatial Efficient Entropic Differencing for Quality Assessment (SpEED-QA) [54] | 2017 | RR | Video |
| Blind Image Quality Index (BIQI) [55] | 2010 | NR | Images - Frames |
| Blind/Referenceless Image Spatial QUality Evaluator (BRISQUE) [56] | 2012 | NR | Images - Frames |
| Naturalness Image Quality Evaluator (NIQE) [57] | 2013 | NR | Images - Frames |
| HDR Image GRADient based Evaluator (HIGRADE) [58] | 2017 | NR | HDR images |

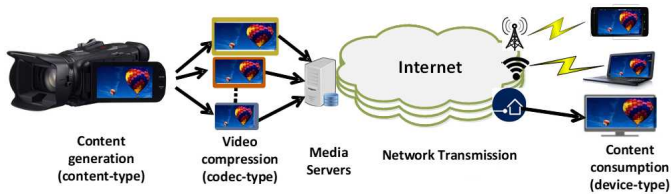

Fig. 4: Multimedia streaming chain.

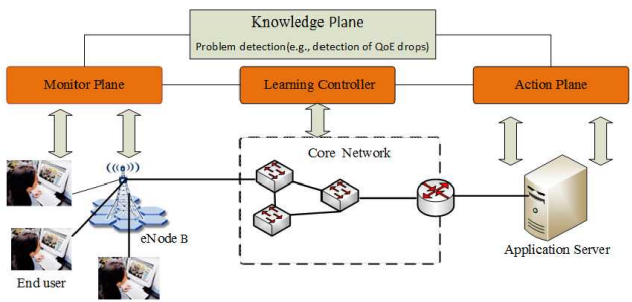

Fig. 5: QoE-driven cross-layer optimization strategy for wireless video transmission.

A QoE-driven service adaptation and network optimization mechanism using resource allocation was proposed in [62] based on the 3GPP IP Multimedia Subsystem architecture (IMS). Given the available network resources and operator's QoE service policy, the aim was to maximize the total utility of the active sessions and calculate the optimal service configurations using the proposed QoS/QoE matching and optimization functions [63]. The QoE-optimization process involves the calculation of the agreed service profile which contains various service level parameters (e.g., frame rate and codec type). The optimization function collects the necessary input data and invokes the optimization engine that runs the QoE optimization algorithms. The final QoE is then estimated based on the negotiated and agreed service profiles between end-users and service providers. Further solutions for 3GPP QoE control mechanisms have been proposed and studied extensively in [64] where a QoE estimation function collects a set of QoE Influence Factors (IFs) and runs at the application server. From the network operators' perspective, QoE-driven optimization involves network resource management mechanisms. Such approaches are typically based on the quality-related information collected from the QoE monitoring and measurements process to provide quality assurance and service control in the network.

QoE-driven optimization has been used for resource management in different networks (e.g., in mobile networks or MESH networks). For example, some of the techniques used include a cross-layer QoE-driven admission control and resource allocation [65] for adaptive multimedia services in LTE networks [66]. Fig. 6 shows a QoE-driven resource allocation scheme for multiple users to access different contents in a wireless environment, as proposed by Thakolsri *et al.* [67]. A utility maximization scheme is applied such that the end-users' perceived quality is improved through rate adaptation and network resources allocation mechanisms. The QoE optimizer in the core network is used as controller and a downlink resource allocator for video rate adaptation. Staehle *et al.* [68] propose *Aquarema*, a network resource management mechanism which improves the end-users' QoE in all types of applications for any kind of network. Using *Aquarema*, QoE degradation is avoided through network management tools and the interaction of application monitoring tools, running at the clients' side.

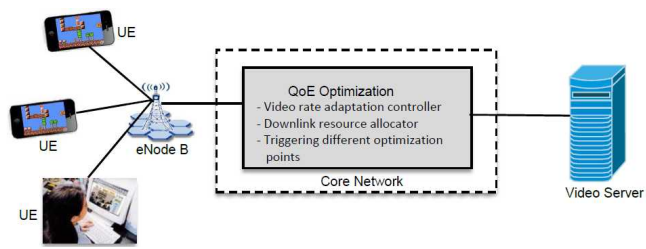

Fig. 6: QoE optimization of multimedia services in access networks.

Wamser *et al.* [69] introduce a YouTube Performance Monitoring Application (YoMoApp), an Android application that can monitor passively KPIs of YouTube adaptive video streaming on end-user smartphones. The monitoring tool runs at the client side to provide a quantification of status for the running applications and enables to predict the end-users QoE. Also, a network advisor is used for triggering different resource management tools during QoE degradation. Such interaction is made possible by using the *YoMo* [70] tool which continually monitors the amount of playtime buffered by the YouTube player.

An autonomic architecture that can optimize the end-users' QoE in the context of multimedia access networks, as shown in Fig. 5, is proposed in [71]. The authors use the monitoring plane to monitor the network and utilize the knowledge plane to analyze and determine the ideal QoE actions. The action plane is used to enforce all of these actions into the network. Two reasoners are applied in the knowledge plane to optimize the end-users' video QoE. The first one is an analytic reasoner while the second reasoner utilizes the feed-forward neural network consisting of one hidden layer and five hidden neurons. Both reasoners are utilized to improve the video quality by reducing the packet loss on a network link as well as the switching impact caused by changing video bitrate [71].

More intelligence in the neural network reasoner can be added by incorporating online learning behaviours such as Q-Learning. In such a case, the learning controller implementing Q-learning behaviour should be capable of finding different QoE optimization actions on a particular service until a better service quality is achieved [72]. However, in another situation and system, other learning/optimization models, e.g., a Convolutional Neural Network (CNN), might be better for QoE maximization [73].

A joint optimization approach of network resource allocation and video quality adaptation that fairly maximizes video clients' QoE is given in [74]. Minimizing energy consumption, especially in the context of mobile services, has been another important considered objective when optimizing the end user's QoE [75]. Tao *et al.* [76] propose an energy-efficient video QoE optimization solution for DASH over wireless networks. The proposed scheme allocates network resource efficiently and make optimal bitrate selection to clients in order to achieve a QoE maximization. Bouten *et al.* in [77] and [78] present in-network quality optimization agents, which monitor the available throughput using sampling-based measurement techniques. That way, the quality of each client is optimized based on a HAS QoE metric.

Triki *et al.* [79] propose a dynamic closed-loop QoE optimization for video adaptation and delivery while Xu *et al.* [80] investigate the buffer starvation of video streaming services. The authors exploit the trade-off between startup/rebuffering delay and starvation to optimize the end-users' QoE of video streaming services. Zhao *et al.* [81] propose a cooperative transmission strategy for video transmission in small-cell networks that reduces video freezes and improves the QoE. The system performance is significantly improved when there are many active users in the network. In that aspect, the greedy algorithm transmits the video-file segments using distributed caching. Furthermore, in [82], the authors propose a central

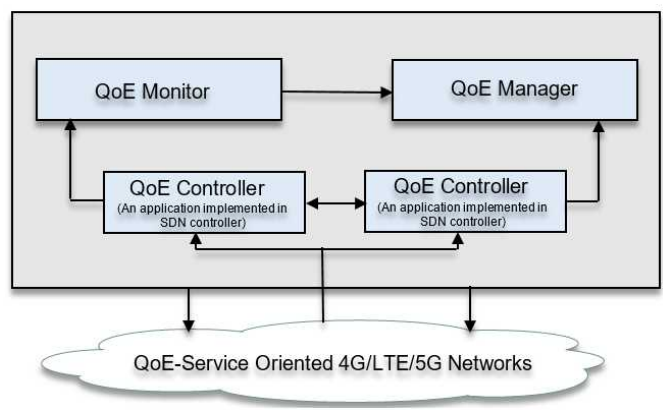

Fig. 7: QoE-based management in mobile cellular networks [82].

QoE management entity to monitor and manage the QoE level of the end-user. The proposed framework, shown in Fig. 7, can collect QoE-related inputs and apply QoE-driven network management decisions. It consists of three components, namely the QoE monitor, QoE manager and QoE controller. The QoE controller is the interface which synchronize communication exchange between the central QoE management entity and the underlying network. It collects the required data and provides them as inputs to QoE monitor and QoE manager. The QoE monitor is used to estimate the QoE per-flow and report this to the QoE manager. It performs traffic classification using statistical analysis and different built-in "QoE models" based on traffic/service type. The QoE manager is responsible to manage the customer's watching experience and the network. It uses inputs from the QoE controller regarding the state of the network to estimate the end-users' QoE. Two or more QoE controllers can be used to improve the end-users' QoE, increase reliability and availability as well as avoid service interruptions during video streaming. More recently, Khan and Martini [83] presented a solution to reduce cross-layer signaling for QoE-driven optimization of scalable video streaming over LTE wireless systems, as also presented in [84],[85].

In order to optimize QoE, Latré and Turck utilize traffic-flow adaptation, admission control, or video rate adaptation mechanisms to achieve a QoE control and management of multimedia services as shown in Fig. 8. The traffic-flow adaptation is used to modify the network delivery of a traffic flow by using redundant data that accompany the application data. The admission control admits or blocks new connections in order to avoid network congestion. The video rate adaptation changes the video quality level as part of the multimedia service transmitted from server to clients. Although the authors in [86] show that these techniques can improve the end-users' QoE, they introduce large deployment complexity because of exchanging information between layers.

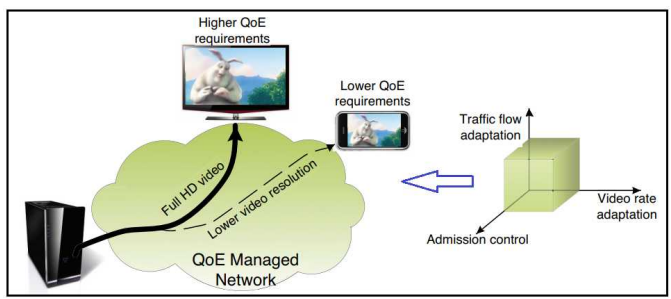

Fig. 8: QoE in multimedia network. QoE is automatically managed using three different axes, namely, (a) traffic-flow adaptation, (b) admission control, and (c) video rate adaptation [86].

*E. Summary*

QoE- based modelling, monitoring, control and management has become topics of high interest in the multimedia communication research community for the QoE-based user-centric service management of the streaming services. The QoE-aware management of multimedia streaming services consists of three major components: *1) QoE modeling and assessment*– that includes the creation of predictive mathematical models (QoE model) from the data related to KQIs; *2) QoE monitoring and measurement*– that includes the monitoring of the KQIs of the multimedia streaming service and measurement of the QoE from the predictive QoE models; *3) QoE optimization and control*– mainly to perform optimization of the resources (network resources, content delivery network or client side adaptation) through control actions based on the QoE measurements.

As stated before, QoE optimization can be performed at various points using different QoE mechanisms, for example at the base stations within the access networks (e.g., using optimal network and radio resource allocation) [87]–[91]. It can be also performed by conducting adjustments of servers in the service/application [71], [91]–[93], applying QoE-based policy management within the core networks [67], [68], [90], [91], or an optimized handover decision [91], on end-user device (e.g., battery consumption) [94]. In addition, QoE optimization can be performed at different levels ranging from link to application-layer [95], [64], or using a common cross-layer approach [67], [68], [90], [91], [83], [96].

## III. MULTIMEDIA STREAMING SERVICES OVER THE INTERNET

With the growth of streaming video traffic, it has become imperative to exploit various factors along the multimedia delivery chain to optimize the video service delivery by considering the end users' QoE. This section presents a comprehensive discussion on HTTP adaptive streaming solutions over the Internet. In particular, it presents the server and client side optimization of video streaming, including Dynamic Adaptive Streaming over HTTP (DASH) as well as Server and Network Assisted DASH (SAND). Furthermore, multimedia delivery chain and service management issues are discussed with a focus on OTTP, ISPs, CDNs, transit providers and IXPs.

*A. HTTP Adaptive Streaming (HAS) Solutions*

The majority of Internet video traffic today is delivered via HAS [6]. The advantages of HAS include: (a) providing a reliable transmission, (b) cache infrastructure reuse capability, and (c) enabling firewall traversal. Based on these benefits, HAS has been adopted commercially as shown in Table IV by Microsoft Smooth Streaming, Adobe Dynamic Streaming, Netflix and Apple HTTP Live Streaming for multimedia streaming service [102]. HAS has been widely used in OTTP video services such as Netflix and YouTube [103] as the de-facto standard for adaptive streaming solutions. The underlying logic is common in all these implementations with some differences in the manifest file, recommended segment size, etc. (see Table IV). HAS solutions use reliable delivery mechanisms such as TCP and very recently Quick UDP Internet Connections (QUIC). However, due to the disparity of these proprietary HAS solutions and the media formats, the 3GPP in close collaboration with MPEG has developed the Dynamic Adaptive Streaming over HTTP (DASH) standard [100].

Fig. 9 illustrates the concept of DASH assuming a video rate adaptation method (e.g., throughput-based, buffer-based, or hybrid) for video streaming. It can be observed that the client, based on its network condition, adapts the quality of the video to provide a smooth streaming experience to the end user. The video is encoded at different representation levels (spatial/temporal/quality) and then divided into chunks (also referred to as segments) of equal duration, which are then stored on a server. When the client makes the first request for the video file, the server sends the corresponding manifest file (e.g., .mpd, .m3u8) which consists of the details about the video file such as duration, segment size, representation levels or codec type [6]. The client then measures/predicts the

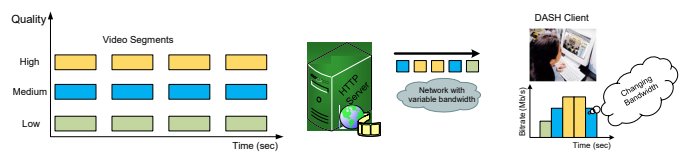

Fig. 9: Dynamic adaptation according to network conditions.

current bandwidth and buffer status and requests the next part of the video segment with an appropriate bitrate. This way, the number and duration of stalling (interruption of playback due to empty playout buffers) could be decreased and the available bandwidth is best possibly utilized [6]. The success of HAS can be attributed to these benefits compared to the traditional

TABLE IV: A COMPARISON OF HTTP ADAPTIVE STREAMING SOLUTIONS.

| HAS Category | Company | Video Codec | Segment Length (sec) | Data Description | Format |
| --- | --- | --- | --- | --- | --- |
| Microsoft Smooth Streaming [97] | Microsoft Corporation | H.264, VC-1 | 2 | Manifest (XML) | fMP4 |
| Apple HTTP Live Streaming (HLS) [98] | Apple Inc. | H.264 | 10 | Playlist file (M3U8) | M2TS, *.ts files |
| Adobe HTTP Dynamic Streaming (HDS) [99] | Adobe Systems Inc. | H.264, VP6 | 2 - 5 | Manifest (F4M) | fMP4 |
| MPEG-DASH [100] | Standard | Any | Not specified | Media Presentation Description (MPD) files (XML) | MP4 or M2TS |
| 3GP-DASH [101] | Standard | H.264 | Not specified | MPD files (XML) | 3GPP File Format |

M2TS = MPEG-2 Transport Stream; fMP4 is a fragmented MP4; MPD = Media Presentation Description

streaming technologies: 1) video service providers can offer multiple quality levels to the end-users' demands by adapting video bitrates; and 2) different QoE-tailored personalized service levels and/or pricing schemes can be offered to customers. It is worth noting that DASH enables both live and on-demand delivery of media streams over HTTP. On-demand video streaming applications can benefit from optimized encoding methods such as multi-pass and variable bitrate encoding. As an example, Netflix proposed a content-aware video encoding optimization [104] strategy where using content information the various optimum resolution-bitrate pairs are determined resulting in better service quality.

Despite the decentralized nature of the HAS principle, there are still some drawbacks especially in the presence of multiple DASH clients that compete for the shared network resources. Some of the HAS related issues include (1) video instability due to bitrate switching [105], (2) network resource under-utilization, and (3) QoE unfairness [106]. These problems remain a serious concern for video content providers and network operators, and they are even more aggravated in the case of heterogeneous environments. Based on these reasons, we discuss next the MPEG-SAND standard where centralized nodes within the network have been proposed recently to enhance the delivery of DASH content [107], [108].

*B. Server And Network assisted DASH (SAND)*

Server and Network assisted DASH (SAND) is an extension of the MPEG-DASH standard that has been recently finalized, [109], [110] to enhance the delivery of DASH content. The SAND specification introduces messages between DASH client and network elements or between various network elements. SAND messages improve the streaming session by providing information about real-time operational characteristics of networks, servers, proxies, caches, CDNs as well as DASH client's performance [109]. SAND has been specifically designed to address: (1) content-awareness and QoE-service-awareness through server/network assistance, (2) analytics and monitoring of DASH-based services, (3) unidirectional/bidirectional, point-to-point/multipoint communication with and without a session (management) between servers/CDNs and DASH clients.

The SAND architecture, shown in Fig. 10, consists of three categories of elements, namely, the DASH clients and DASH-Aware Network Elements (DANE). It also consists of regular network elements (which are DASH unaware) that resides on the path between the origin server and DASH clients (e.g., transparent caches). DANE communicates with the DASH clients while having minimum intelligence about DASH. For example, DANE nodes may be aware that the delivered objects are DASH-formatted objects such as the *mpd* or DASH segments. This way, they can prioritize, parse or even modify such objects. The MPEG-SAND standard reference architecture, shown in Fig. 10, also defines three categories of SAND messages, namely: (1) the Parameters Enhancing Reception (PER) messages which are sent from DANEs to DASH clients for enhancing and improving their video quality adaptation, (2) Parameters Enhancing Delivery (PED) messages which are exchanged between DANEs, and (3) metrics and status messages which are sent from DASH clients to DANEs.

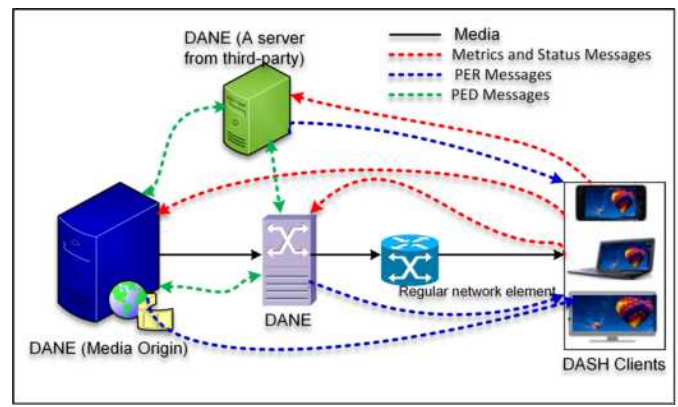

Fig. 10: MPEG-SAND architecture. DANEs can communicate among each other and with the DASH clients to optimize the end-to-end video quality delivery (adapted from [109]).

DANE nodes become aware of the status of DASH clients using SAND status messages. For example, QoE metrics reported by DASH clients to the network can be used for monitoring purposes and to simplify video data rate optimization implementations. A DANE node can use PER messages to inform the DASH clients about the available network bandwidth. They can also inform clients about already cached segments by the DANE so that the clients can request these segments based on their device capabilities. Moreover, information about the streamed video to a particular network delivery element/node can be communicated by the server using a PED message. Generally speaking, a third-party server that receives messages concerning metrics from DASH clients and sends SAND messages to the clients is considered as a DANE element. It is important to note that the MPEG-SAND messages are delivered over HTTP using the XML format and follow a specific syntax as defined by the standard in [109]. Compared to the client or server-based solutions, the MPEG-SAND standard represents an essential enabler for solutions that need collaboration between service/application and network providers. However, this requires a modification of the network nodes/elements to provide a set of messages that can be exchanged by the network, servers, and DASH clients during video quality delivery optimization [110]. For the practical and scalable implementation of the SAND approach, SDN can be used to provide a centralized control element. Efficient QoE-driven application and QoE-aware network management strategies using SDN and NFV appear to be vital solutions to guarantee the end users' QoE at a minimum cost.

*C. Multimedia Delivery Chain and Service Management Issues*

The multimedia service delivery chain involves several entities which include OTTP, ISPs, CDNs, transit providers and IXPs, where each entity has a different role depending on their mutual agreements [111]. Currently, there are ongoing QoS based collaborations between the OTTP and ISPs concerning peering agreements and OTTP's surrogate server (SS) hosting in ISP's network depending on the agreement [112]. For example, Google is providing peering connections for ISPs through IXPs, and also Surrogate Servers (SSs) are being provided under the Google Global Cache (GCC) program[2]. Similarly, Netflix is providing private peering connections and SSs to the collaborating ISPs under the Netflix Open Connect program[3]. However, in the collaboration scheme mentioned above, the ISPs only have multi-hop peering connections with the Autonomous System (AS)[4]. The ISPs can host the SSs in the network edges, but the OTTP manage the distribution of the content in the SSs and the assignment of the appropriate SS to the end users.

The multimedia content delivery operations on the OTTP side can be broadly divided into two categories: 1) Client request redirection and; 2) Content distribution and replication [115]. On the ISP side, the network management operations are mainly composed of the traffic engineering abstractions, which include optimal routing strategies, prioritization, and dynamic resource allocation. The following subsections provide the details of how OTTP and ISP may assist each other in collaborative QoE service management.

*1) Client Request Redirection and Optimal Route:* The request redirection is a crucial operation in multimedia content delivery services that redirect client requests to the nearby CDN or SS to reduce the content retrieval time. The request redirection mechanism can be based either on Domain Name System (DNS) unicast or on a combination of DNS and anycast for assignment of the most nearby server to the clients [116]. To control the namespace, the OTT/CDN providers have many authoritative nameservers. The namespace is also divided into subnamespaces for load balancing to resolve the request. The content to be delivered is assigned to a nameserver by the Uniform Resource Locator (URL) where the prefix represents the nameserver, and the suffix identifies the multimedia content.

However, the DNS based request redirection relies only on the link layer-information for CDN/SS assignment to the clients, which does not allow it to react to the congestion in the ISP's network. In this case, the OTTP cannot deliver adequate quality to the end users while the ISP has to transport data through a non-optimal path since no information is being shared between the OTTP and ISPs. In the worst case, the client is assigned to a CDN outside the ISPs network which will not only decrease the quality but also results in a high cost for the ISP and an increased network load in the ISP's core network. In this case, SDN can provide an opportunity to the ISPs to assist OTTP for the optimal selection of a CDN/SS considering their network [116].

*2) Content Distribution/Replication and Cache Miss Handling:* Content distribution and replication into CDNs and SSs from the original server can be performed in two ways: *Hierarchical* and *Flat*. In the Hierarchical organization, the content distribution/replication is performed in a multi-tree like fashion where HTTP redirect is used for moving the content up and down in the tree between original server, CDNs, and SSs. In this case, information of the content location can be embedded into the overlay to avoid cache misses. For example, Google has implemented a hierarchical distribution of content for YouTube and other Google services [117], [118]. On the other hand, the OTTP services utilizing Akamai CDNs deliver the content based on the flat organization where cache misses is avoided by fetching the content directly from the original server [119]. However, if the content is not available in the last mile SS, the client is served by the CDN/SS located far from the client's location. This can lead to cache misses and higher end-to-end delay which can decrease the delivered quality.

Moreover, all SSs can serve many clients that request a video content simultaneously. In order to balance the load, some of the clients are served from the last mile servers [120]. Since the OTTP's SSs are physically located in the ISP's network, the SSs cannot be shifted in case of the appearance of a flash crowd or congestion in the network. However, NFV can provide an opportunity for the secure deployment and movement of Virtual Surrogate Servers (VSSs) as VNF to avoid the quality degradation cases mentioned above. Nonetheless, this calls for collaboration among OTTP and ISPs, which indeed requires an interface for the exchange of information among the two entities [2]. The ISPs have to provide OTTP with the access to the VNF infrastructure for content replication/distribution [121].

*D. Summary*

Multimedia streaming services by OTTP are using HTTP adaptive video streaming based on the MPEG-DASH standard introduced in section III-A. While the MPEG-DASH standard can provide excellent video quality to the end-users, there are some issues when multiple clients are competing for the shared limited resources leading to QoE unfairness, video instability and under-utilization of the network resources [105], [106]. To overcome these challenges, the MPEG-SAND standard has been proposed recently with the aim to enhance the delivery of DASH contents, thanks to its ability to provide information of the network, servers, and the streaming DASH clients in real-time. DASH clients can announce information to DANEs regarding their required operating video data rate, a set of segments and video quality. DANEs can send information to the DASH clients about the video segment and the network throughput availability, as well as caching status of the segments. From the multimedia delivery chain and service management perspective, OTTP and ISPs may assist each other in collaborative QoE service management paradigm using a) client request, redirection and optimal route, b) content distribution and replication.

---

[2]https://peering.google.com/
[3]https://openconnect.netflix.com
[4]The AS represents a connected group of one or more blocks of IP addresses called IP prefixes whose routing policies are under common administrative control [113,114].

## IV. NETWORK SOFTWARIZATION AND VIRTUALIZATION: THE PROMISE OF SDN AND NFV IN FUTURE NETWORKS

Network softwarization [122] and virtualization using SDN and NFV is expected to impact several aspects in the development and deployment of network services, such as CDN and video accelerators [123], [124]. Before discussing approaches for QoE management using SDN and NFV in Section V, we introduce the reader to the recent advancements in SDN and NFV as an important technology for management and orchestration of resources in future networks. In that respect, Section IV-A discusses the SDN controller design and practical issues. This is followed by Section IV-B, which describes the NFV Management and Orchestration (NFV MANO) framework, NFV use cases and application scenarios as proposed by the ETSI. A description of the relevant ongoing research projects and standardization activities from different bodies/consortia that are pushing forward the adoption of SDN and NFV for multimedia delivery is given as well.

### A. Software Defined Networking (SDN)

SDN [12] is an approach that brings intelligence and flexibility in programmable networks to allow for orchestrating and controlling the running applications and services in a more fine-grained and network-wide manner with respect to past approaches [125]. The Open Network Foundation (ONF) [126] defines SDN as "*the physical separation of the network control plane from the forwarding plane, and where a control plane controls several devices*". Notable advantages of the SDN architecture include enhanced network programmability, centralized control and management, increased network flexibility and reliability, data flow optimization [126]. As shown in Fig. 11, SDN creates a virtualized control plane that can enforce intelligent management decisions among network functions bridging the gap between service provisioning and QoE management. The QoE control and management plane of an SDN network can be implemented as a software that runs on commodity hardware devices. With SDN, the network control becomes directly programmable using standardized Southbound Interfaces (SBI) such as OpFlex [127], FoRCES [128], and OpenFlow [129]. The forwarding plane of SDN can be implemented on a specific commodity server [130] such as VMware's NSX platform [131] which consists of a controller and a virtual switch (vSwitch).

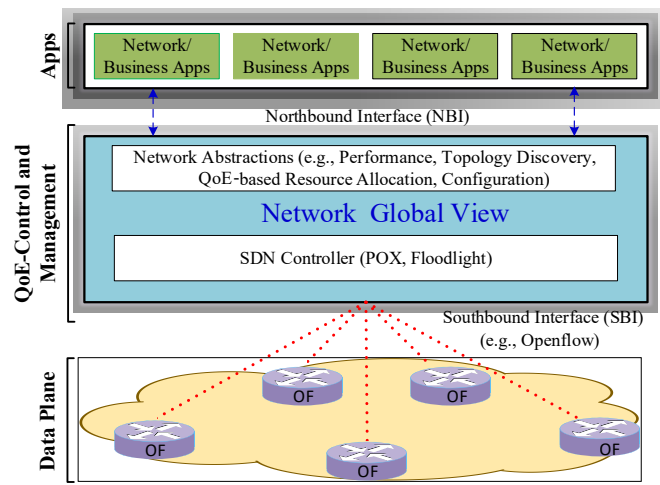

Fig. 11: Main layers of the SDN architecture.

*1) SDN Controller Design and Implementation:* The design and implementation choice of an SDN controller significantly affects the overall performance of SDN-based networks [12]. Key design and architectural choices for the controller are: *centralized* or *distributed*. A centralized SDN controller is best suited to manage small networks or a single domain. However, network management using the centralized controller can face scalability and reliability problems [132]. The controller itself can be a point of failure which can negatively affect the performance of the network and the end-users' QoE. SDN controllers such as Floodlight [133], Beacon [134], and Ryu NOS [135] have been designed to fulfill the needs of managing network resources in data centers, carrier-grade networks, and cloud infrastructures. It is important to note that a centralized controller is not enough for large-scale network management. Scalability and reliability in large-scale networks that span multiple control domains can be achieved by distributed SDN control architectures such as Hyperflow [136], HP VAN SDN [137], DISCO [138], and ONOS [139]. In essence, a distributed SDN controller implementation can be either a set of elements distributed physically or a centralized cluster of nodes [12]. It is worth mentioning that a distributed SDN controller reduces the network partitioning[5] problems and improves the scalability and resilience of the control plane in SDN [140].

*2) Standardization Activities in the SDN Domain:* Driven by the potential of SDN towards simplifying the QoE control and management of multimedia services in the Next Generation Networks (NGN), major standard providers such as the ONF, 3GPP, ETSI, IEEE, and the Internet Engineering Task Force (IETF) are working on different standardization activities of SDN. For example, the ONF [126] introduced the SDN architecture and developed the OpenFlow standard that is now used for commercial deployments. The IETF working group FORCES [141] defines interfaces and protocols for network infrastructure abstraction, the separation of forwarding and centralized network control. The SDNRG [142] investigates technical aspects of SDN including solutions for scalability, abstractions, programming languages and paradigms particularly useful in the context of SDN. The IEEE standards such as IEEE P1915.1 [143], IEEE P1916.1 [144], and IEEE P1917.1 [145] specify requirements, frameworks, and models on the aspects of security, performance, and reliability of SDN.

---
[5]A network partitioning refers to network decomposition into relatively independent subnets for their separate optimization as well as network split due to the failure of network devices

ITU-T SDN standardization activities span across several Study Groups (SGs). For example, SG11 investigates signaling requirements and protocols of SDN for Wireless Access Network (WAN) services and bandwidth adjustment on the broadband network gateway. SG13 specifies use-cases, requirements, and architecture of SDN for NGN. The Metro Ethernet Forum (MEF) [146] has now introduced SDN technologies to the carrier Ethernet services. Continually evolving, the MEF facilitates industry-neutral implementation environments for service orchestration and L2-L7 connectivity services based on open source SDN and NFV. The open source software community such as OpenStack [147], CloudStack [148] and OpenDaylight [149] are accelerating the adoption and fostering innovation by developing the basic building blocks of SDN.

### B. Network Function Virtualization (NFV)

NFV [13] is the decoupling of physical network equipment from the network functions such as Firewalls and Deep Packet Inspection (DPI) that run on them. NFV envisages the instantiation of Virtual Network Functions (VNFs) on commodity hardware. With NFV, Network Functions (NFs) can be quickly deployed and dynamically allocated [150]. Also, network resources can be efficiently allocated for these VNFs to achieve Service Function Chaining (SFC)[6]. For ISPs, NFV promises to provide the needed flexibility that would enable them to support new network services faster and cheaper. It also enable them to realize better service agility and reduce their CAPital EXpenditure (CAPEX) and OPerational EXpenditure (OPEX) through lower-cost flexible network infrastructures. NFV also aims to decrease the deployment time of new network services to market and support changing business requirements. To achieve the above benefits, NFV brings three differences in the network services provisioning as compared to the traditional practice [13], [152], by (a) decoupling software from hardware platform, (b) providing greater flexibility for NFs deployment, and (c) enabling dynamic network operation and service provisioning. The question is whether the NFV design considerations can meet all the technical performance requirements needed by Telco Cloud or service providers. The NFV Management and Orchestration (MANO) framework as proposed by the European Telecommunication Standard Institute (ETSI) is discussed next.

*1) NFV Management and Orchestration (NFV MANO) Framework:* The NFV concept in operator infrastructures [153] was first explored by the ETSI, mostly to address the challenges towards flexible and agile services and to create a platform for future network monetization. Since then, the NFV reference architecture, as shown in Fig. 12 was proposed [154], followed by a proof of concept (PoC) [155]. The ETSI MANO framework consists of functional blocks which can be grouped into the following categories: the NFV Infrastructure (NFVI), NFV Orchestrator (NFVO), Network Management System (NMS) and VNFs and Services. These entities or blocks are connected using reference points[7].

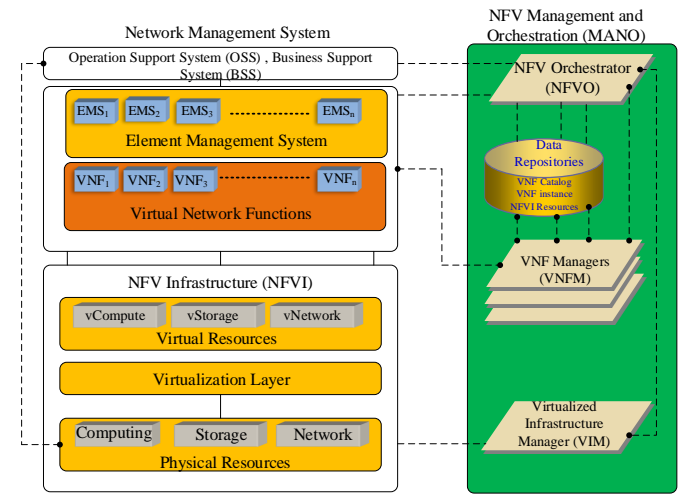

Fig. 12: The ETSI NFV MANO framework.

The NFVI forms an environment consisting of both physical and software/virtual resources. While physical resources indicate the computing hardware such as storage and network resources, virtual resources are the abstractions of the computing, storage and network resources achieved through a virtualization layer (e.g., based on a hypervisor which is a typical solution for VNFs deployment today). NMS is a critical aspect of the overall operation of both VNFs and physical resources. It potentially deals with functions related to network management such as fault management, security, configuration and performance management. The fault management provides a key role for QoE assurance by making sure that network failures or problems are recovered before users are disconnected from the network services.

The NFVO performs orchestration and lifecycle management of physical and software resources that support the virtualized infrastructure. The NFVO also performs global resource management, network service instantiation, validation, and authorization of NFVI resource requests. The VNFM is responsible for lifecycle management of VNF instances, overall coordination for configuration and event reporting between NFVI. A VNFM can manage a single or multiple VNF instances of the same or different types. The VIM controls and manages NFVI physical and virtual resources (vCompute, vStorage, and vNetwork resources) within operator's infrastructure domain. The NFV MANO contains a database that keeps the VNF catalog, VNF instances, NFVI resources, and network services catalog. The NFVI resource

---

[6]Service Function Chaining (SFC) is an ordered list of general service functions that should be applied to a packet and flows selected as a result of classification [151].

[7]A reference point defines a point where two communicating functional entities or blocks are connected.

repository holds information of all allocated or available NFVI resources.

*2) NFV Use Cases, Application Scenarios and Implementation:* As NFV became important, many use cases and application scenarios from both the academia and industry have been designed and implemented. Most of these use cases are based on those defined by the ETSI [156]. Some of the use cases from the academia include those pertaining to QoE-based multipath routing [157] or virtual function implementation for broadband remote access server [158], deep packet inspection [159], radio access network (RAN) [160]–[162], customer premises equipment (CPE) [163]–[165], evolved packet core (EPC) [166]–[168] and QoE monitoring through virtualized probes [60]. From the industry, key NFV implementations and products include the CloudNFV [169], Huawei NFV Open Lab [170], HP OpenNFV [171], Intel Open Network Platform (Intel ONP) [172], Cisco Open Network Strategy [173], Alcatel-Lucent CloudBand [174], [175], Broadcom Open NFV [176] and F5 Software Defined Application Services (F5 DAS) [13]. It is worth noting that all of the existing NFV implementations and platforms are focusing on open source and orchestration of operator's end-to-end service over NFVI supported by SDN and cloud technologies.

*3) NFV Standardization Activities:* Current NFV standardization activities span across many domains. The NFV Research Group (NFVRG) [177] of the Internet Research Task Force (IRTF) is active in developing new virtualized architectures with capabilities to provide support for NFs. The IETF SFC Working Group (IETF SFC WG) [178] has put efforts to develop SFC architectural building blocks that address mechanisms for autonomic specification, instantiation of NFV instances and steering data traffic through service functions. The ETSI's Industry Specification Group for NFV (ETSI NFV ISG) [179] provides use cases for NFV requirements, architectural framework [154], [153], interfaces and abstractions for NFVI, NFV security [180], performance and resiliency [181].

As a follow-on activity, the Alliance for Telecommunications Industry Solutions (ATIS) NFV Forum (ATIS NFV Forum) [182] is devoted to identify, define and prioritize use cases for collaboration among service providers where NFV capabilities are required to generate new value. In partnership with the ETSI, the Broadband Forum [183] has been working on introducing NFV into the Multi-Service Broadband Network (MSBN) and has established a virtualized platform for virtual business gateway and flexible service chaining [184]. Other bodies such as the Open Virtualization Format (OVF) and the ITU-T SG13 have been working towards defining the functional requirements and architecture of the network virtualization for NGN as well as the portability and deployment of both virtual and physical machines across multiple platforms. Along with the above standardization activities, many collaborative projects pushing the NFV implementations include the Open Platform for NFV (OPNFV) [185], the Zero-time Orchestration, Operations and Management (ZOOM) [186] and OpenMANO [187], Unifying compute and network virtualization (UNIFY) [188], [189] among others.

*C. Summary*

Standardization activities have been focused on identifying how network services and the associated resources that are implemented, according to an SDN architecture, might be integrated within the NFV architectural framework [190]. It is worth noting that both SDN and NFV seek to create a future software-based networking solution that offers flexible and automated network connectivity and QoE provisioning to the end-users. For example, while SDN decouples the control plane from the data/packet forwarding plane, the NFV decouples NFs from dedicated hardware devices. Although the two (SDN and NFV) have a lot in common, yet their main difference is that SDN requires a new network platform where the control and data forwarding planes are decoupled. This is not the case with NFV which can run on legacy networks since NFs can reside on commodity servers.

## V. QoE MANAGEMENT USING SDN AND NFV

Since the MPEG-DASH standard was defined, a lot of efforts from both the academia and the industry have investigated how to improve the end-users' QoE. Many studies have been focusing on enhancing the QoE by optimizing the video quality adaptation algorithms on the client and/or the server-side [6], [191]. In recent years, the attention has been towards developing approaches that fully optimize the video quality [192], [107] and maximize the QoE while ensuring fairness among users [193]–[195]. As of today, many SDN/NFV based solutions have been proposed, implemented and tested to see the value and benefits offered by these technologies. Recent efforts include a QoE-aware bandwidth broker [196] and a rate-guided QoE-aware SDN-APP [197].

In this section, we focus on different research areas. Server and network-assisted optimization approaches using SDN are analysed in Section V-A, whereas QoE-fairness and personalized QoE-centric control are reviewed in Section V-B. Section V-C describes a detailed QoE-centric routing while Section V-D presents QoE-aware approaches based on the Multipath Transmission Control Protocol (MPTCP) and Segment Routing (SR). The SDN/NFV-based collaborative service QoE management by OTTP and ISPs is described in section V-E. Finally, Section V-F analyzes proposals that employ "*full adoption*" of SDN and NFV during QoE-optimization. The final subsection summarizes these QoE-centric management approaches with the help of Table V.

*A. Server and Network-assisted Optimization Approaches using SDN*

In order to improve the efficiency of HTTP video streaming, SAND provides standardized messages between DASH clients and network elements. It also enables service providers and operators to improve network bandwidth utilization and enhance the end-users' video streaming experience. SAND enables HAS based streaming solutions to provide a fair distribution of bandwidth among users. With SAND, bandwidth reservation and bitrate guidance strategies have attracted much attention in the multimedia streaming community [7], [107] [198], [196]. Most research works following this approach share a similar architecture, as shown in Fig 13. Relevant quality parameters from both the network and the client side can be collected using a centralized node. The client-based information collected may include, for example, the device type [7], the bitrate of the video [199], the buffer occupancy level [198] and the screen size. The network QoE-related parameter measurements include the number of streaming DASH clients and the available bandwidth. That way, the centralized node in an SDN-assisted environment has a comprehensive view of the streaming multimedia service. It is hence possible to select the best bitrate for each client to optimize an objective function that includes models of the users' QoE. For the practical and scalable implementation of this approach, SDN can be used to provide a centralized control element.

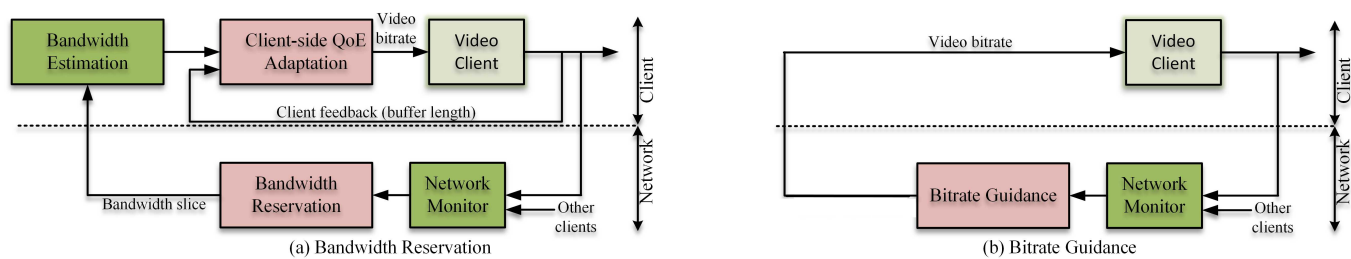

Fig. 13: QoE-based network-assisted approaches for adaptive video streaming. (a) A specific bandwidth for each client is enforced by the network when bandwidth reservation is used, and (b) the network provides an explicit bitrate to the clients using bitrate guidance [107].

Authors in [107] compare the performance of bitrate guidance and bandwidth reservation techniques. The optimal bitrates for DASH clients are computed by the SDN controller to achieve video quality fairness, which is calculated based on the SSIM index. When *bandwidth reservation* is used, the bandwidth slice is assigned to clients with similar optimal bitrates. As shown in Fig 13 (a), two control loops, *inner and outer control* loops are used. Based on the video client feedback and bandwidth estimates, the inner control loop running at the client side selects the video bitrate. The outer control loop is executed in the network and sets the bandwidth slice. In the bitrate guidance scenario shown in Fig 13 (b), the optimal bitrates are computed by a centralized algorithm running in a network element. The video bitrates are then communicated to the DASH client that downloads the corresponding video segment.

Following the same approach, authors in [200] propose a network-level QoS mechanism to enforce application QoE-aware fair resource sharing among competing clients. Using the QoE model that depends on screen size and resolution, the network controller selects the best video bitrate to maximize the QoE for each client. A Video Home Shaper is designed to monitor outbound HTTP requests and capture those that identify Netflix and YouTube sessions initiated by clients connected to the router. The QoE-fair bandwidth for all active streaming sessions is computed and recorded by the session manager. The bandwidth manager allocates the guaranteed minimum bandwidth to each video stream and enforces QoE-fair allocations calculated by the session manager.

An SDN-assisted Adaptive Bitrate Streaming (SABR) is proposed in [108]. The available bandwidth per link and network cache contents are the information used by SABR to guide the QoE maximization of the clients. SABR provides the DASH client with monitoring information (bandwidth estimates and cache occupancy) through a REST API. SABR uses dynamic SDN routing to provide clients with the ability to connect to the desired cache. It is worthwhile emphasizing that not only does SABR significantly improve the quality (e.g., through the overall video bitrate) at the client side but also reduces the server load ratio and provides higher network utilization.

A QoE-aware video segment selection and caching approach in the context of HAS (*QoE-SDN APP*) is proposed recently by Liotou *et.al* [197]. Taking into account the user movements and network conditions, the desired users' QoE enhancement from Video Service Providers (VSPs) is achieved by using network-aware video segment selection, efficient encoding rate and caching strategies that reduce stalling events. The core logic of the QoE-SDN APP is the QoE assessment module which performs the following tasks (a) determine and recommend the encoding rate and caching strategy that should be adopted by VSP taking into account the future network load and user mobility, and (b) determine the QoE per application using application-specific KPIs such as stalling events. VSP-QoE Control Agent is another module within the SDN controller that provides feedback to the VSP regarding the control capabilities related to the data plane. It allows VSPs to collaborate with the underlying Mobile Network Operators' (MNOs) infrastructure. That way, VSPs can quickly enhance their distribution procedures and video segment encoding by using network feedback exposed by the MNOs.

Bentaleb *et al.* [196] propose a QoE-aware Bandwidth Management Solution (BMS) for HAS flows in SDN-enabled Hybrid Fiber Coax (HFC) access networks. To improve the end users' QoE among heterogeneous competing clients (e.g., smartphones, tablets, TVs) that have different device capabilities (e.g., screen resolution, CPU and memory), the BMS performs bandwidth allocation, slicing and monitoring for one or a group of video streaming sessions [196]. The main entity of the Bandwidth Management Application (BMA) which is implemented in the application plane is the Viewer QoE Optimizer (VQO). The VQO is responsible for computing the optimal joint video presentation decisions that optimize the QoE of the viewer. The VQO performs three tasks. It first determines the optimization granularity for a single session (e.g., unicast VoD or OTTP services) or group (broadcast/multicast services) of video streaming sessions. Then, the VQO uses the concave network utility maximization (NUM) function to formulate the joint representation decision and bandwidth allocation that maximize the viewer QoE. Finally, it uses a collection of models and methods such as the fast Model Predictive Control (fastMPC) [201], the SDN optimization language (SOL) [202] and the optimal online decomposition [203] to solve the overall QoE optimization problem of the viewer.

### B. QoE-Fairness and Personalized QoE-centric Control in SDN

We provide in this subsection a detailed discussion on QoE-fairness and personalized QoE-centric control approaches for adaptive video streaming. Most of the works presented in this subsection utilize the SDN controller to manage and monitor all HAS players, their statuses, device capabilities, requested content, subscription plan types, QoE and buffer levels. In this way, the controller can easily detect player-specific events, such as players joining/leaving the network and starting/stopping the playback. Some of the approaches, for example in [7,198], employ an optimization function that interacts with the controller to dynamically optimize QoE fairness of multiple competing clients by setting the bitrate for each streaming video in the network. Note that, some of the approaches discussed in the previous section V-A, such as [107], address QoE-fairness based on the SAND architecture. This is different from the approaches discussed in this subsection that are only based on the SDN controller to control and manage QoE requirements of the competing clients.

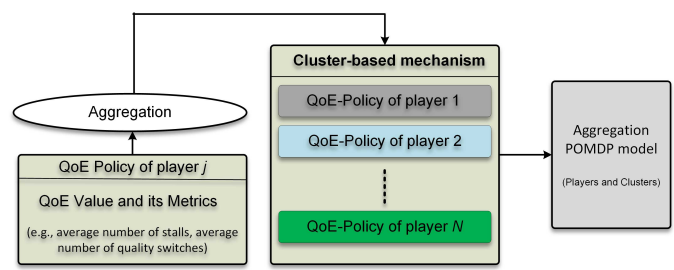

Fig. 14: Per-cluster QoE policy structure and model abstraction (adapted from [106]).

Based on the principles of SDN and Participatory Networking (PANE) [204], [205] proposes a client-Driven Video Delivery (cDVD), a proof-of-concept that provides a client-level API into the network. cDVD relies on a client-driven network-level QoE fairness approach that provides stable video quality by maintaining low re-buffering ratios in an encrypted SDN-assisted environment. Rebuffering ratio is the ratio between the rebuffering duration and the actual time of watching a video content. cDVD at a high level is set to enable one or more DASH clients to interact with network components (e.g., SDN controller) to achieve session-level QoE-fairness. As presented in the cDVD architecture [205], each session can comprise of multiple flows where the controller applies bandwidth control to the routers using a suitable mechanism, such as OpenFlow or direct queue configuration. The controller uses the bandwidth enforcer to limit the rate of each flow and in turn allocates bandwidth to equalize QoE across sessions. This is achieved by applying a QoE-based utility module to determine the fairness of video streams [200].

An OpenFlow-Assisted QoE Fairness Framework (QFF) is introduced by [7] to provide the required user level QoE in multimedia networks. QFF can allocate network resources and ensure that the maximum number of users with the target QoE fairness level is achieved in a heterogeneous environment. In QFF, the OpenFlow protocol allows vendor-agnostic functionality to be implemented for network management and dynamic resource allocation. As such, the status of the network and DASH video streaming sessions are monitored by QFF using the Network Inspector and the MPD Parser module, respectively. The QFF, in turn, dynamically allocates network resources to each client to equitably maximize users' QoE in multimedia networks. The network intelligence of QFF is provided by the Utility Function (UF) and the Optimization Function (OF). The UF offers a model that maps the bitrate of a video to the QoE delivered on a particular client's device. For each video streaming session in the network, the OF is responsible for finding a set of video bitrates that will provide a QoE-fairness level for all DASH clients [7].

SDNDASH, a dynamic network resource allocation and management architecture for HAS systems are proposed by Bentaleb *et al.* [198]. SDNDASH avoids quality instability, unfair bandwidth sharing and network resource under-utilization among competing DASH clients sharing the same bottleneck network link. In this way, the per-client QoE is optimized while reaching the required maximum QoE-fairness level. As an extension to this work, [106] proposes an SDN-based streaming solution called SDNHAS that can assist HAS players in making better QoE adaptation decisions. SDNHAS can optimally implement the target QoE policies for a group of users and allocate the network resources efficiently in the presence of both short and long-term changes in the network. One of the central SDNHAS entities is an optimizer component shown in Fig. 14 that forms a logical network topology to group the HAS players into a set of virtual clusters. A specific data structure for each cluster, called the *per-cluster QoE policy*, is constructed during each video segment being downloaded. To make a fair allocation of bandwidth, each player has its own QoE policy that also includes QoE values and its metrics. A set of players, whose QoE policies belong to the same cluster are aggregated together into a standard per-cluster QoE policy using a simple aggregation method [106]. It is important to note that SDNHAS provides QoE-aware adaptive streaming delivery and intelligent network management that enables a maximum level of user satisfaction among heterogeneous HAS players.

Following the same approach of QoE-fairness and QoE-personalized control, an SDN-based multi-client bandwidth management architecture for HTTP adaptive video streaming that can support up to 75% users at the same QoE level is proposed in [206], while a Q-learning-based dynamic bandwidth allocation strategy to achieve QoE fairness is given in [207]. A user-level fairness model, UFair, which orchestrates network resource allocation between HAS streams to mitigate QoE fluctuations and improve the overall QoE fairness is given in [193]. UFair uses the video quality, switching impact and cost efficiency metrics to measure the users' QoE fairness. Following specific adaptation and fairness criteria of equal bandwidth for every active DASH client, high-quality videos are delivered over a DASH-aware SDN-based architecture in [199]. While heading towards 5G networks, what is exciting and remains to be seen is how QoE-fairness [194] metrics can be integrated as the benchmark for QoE management and optimization in future networks.

### C. QoE-Centric Routing Mechanisms using SDN/NFV

Efficient delivery of video streams with improved QoE can be achieved in SDN using shortest paths, multiple disjoint paths or IP multicast procedures. For example, using the concept of Economic Traffic Management (ETM)[8], a QoE-centric routing approach that utilizes QoE estimation models to maximize the user QoE for multimedia services is presented in [208]. The QoE measurement and collection of QoE IFs values in an SDN network is achieved through the cooperation among various components including clients, media content servers and SDN controllers. The clients and content server report the QoE values to the SDN applications, which in turn use the reported QoE values and IFs to prepare the needed input for the SDN controller. Subject to traffic demands and network constraints, the controller can select a path that can maximize QoE for DASH clients.

A QoE-driven path optimization model (Q-POINT) that maximizes the end users' QoE through the best path calculations for each service flow is given in [209]. In order to negotiate important parameters for video streaming sessions that are to be established, Q-point utilizes the Session Initiation Protocol (SIP) [210] which is assisted by the SIP application server with QoS-QoE mapping functions (for different media types) and Optimization Function (QMOF). For each session, the QMOF calculates a set of configurations that include information such as media codecs, video bitrates, and user preferences. In turn, the SDN controller determines which multimedia flows are to be routed along a particular path in order to maximize the aggregated QoE. The routing decision by the SDN controller is made based on the calculated session and media flow parameters.

A QoE-driven multimedia service optimization and path assignment architecture is proposed in [211]. The main idea is to include various application-level network functions involved in the negotiation and QoE-optimization decision making [212]. A software-defined scalable multimedia multicast streaming ($SDM^2Cast$) [213] is introduced to provide in-network adaptation by adjusting the number of layers, which in turn improves the end-users' QoE. Based on the current network state, $SDM^2Cast$ can flexibly customize network layer multicast paths, thereby allowing SDN implementations to recognize, process and manage media streams through in-network adaptation. An SDN-based video streaming multicast application that performs the calculation of routes and multicast trees after mapping clients to servers with the highest QoE is presented in [214].

---

[8]ETM is an economic incentive that enables a Triple-Win solution for users, service providers and network operators. It aims to reduce cost in the network while providing QoE-enriched services to the end-users.

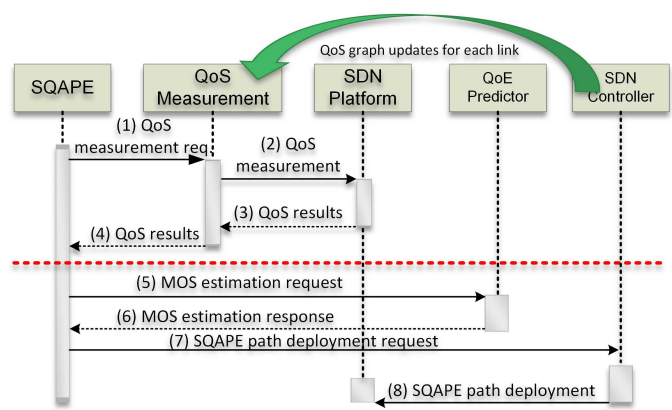

Fig. 15: Scalable QoE-aware path selection in large-scale SDN-based networks.

Athanasopoulos *et al.* [215] proposes a quality-aware path switching scheme, along with a media-aware optimization algorithm that selectively shapes the video traffic according to the access network conditions. Authors in [216], introduce a QoE-aware SDN based video streaming framework that dynamically changes routing paths using Multi-Protocol Label Switching (MPLS) Traffic Engineering (TE) to provide to clients a reliable video watching experience. [217] demonstrates a Scalable QoE-aware Path Selection (SQAPE) scheme for large-scale SDN-based mobile networks. Using a centralized control strategy of SDN, SQAPE provides fine-grained control for per-user QoE-aware path deployments across the network. As shown in Fig. 15, it consists of the *QoS measurement, and QoE predictor* components, which are decoupled from the SDN controller.

As shown in Fig. 15, steps (1) - (4) involve QoS measurements to predict the QoE and the performance of video streaming. After every 60 secs, the network monitoring is performed to measure the QoS based on packet loss, delay, and bandwidth of the link. Such metrics are selected based on previous investigations of QoS to QoE mapping [218], where packet loss and delay metrics were also found to affect the video quality in real-time video streaming services. SQAPE in steps (5) - (8) relies on the QoS-to-QoE mapping function to compute per-path MOS estimation. The estimated MOS, along with a link utilization heuristic is then used to compose the delay metric, which in turn, determines optimized QoE-aware paths to be deployed in SDN-enabled networks. An optimal route for each client is recomputed by [219] each time a new video segment is requested.

The video bitrate requested by a client is sent by a server to the SDN controller. The controller uses this information and the links' status to re-route the traffic to maximize the throughput of the overall link. QoE-aware routing can also be performed using multiple video servers where the client selects dynamically the best server to stream the video [220], [221].

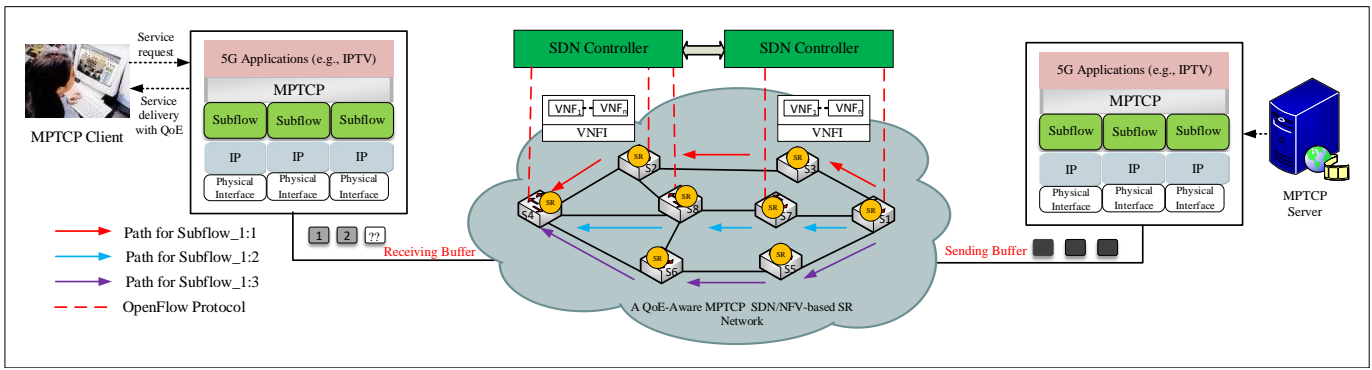

Fig. 16: A video streaming scenario using MPTCP in SDN/NFV-based networks. The black boxes indicate video frames generated as data units by the application while grey boxes are the TCP packets to be received at the client side. One packet is lost and re-transmitted.

Uzakgider et al. [222] propose a learning-based approach that minimizes the packet loss rate, quality changes, and controller cost while adapting the flow routes and video quality. Gangwal et al. [223] propose an Efficient Layer Based Routing Algorithm (ELBA) that exploits the dynamic re-routing capability of SDN, to different stream layers of Scalable Video Codec (SVC) video over distinct paths. ELBA improves network resources utilization and optimizes the video delivery, based on the routing decision of the SDN controller that assigns different routes for each video layer. It is important to note that traffic routing improves QoE by continuously monitoring the performance of the paths that connect the server and clients. However, novel paradigms such as MPTCP and Segment Routing (SR) can facilitate efficient routing and speed up the transfer of large amounts of multimedia applications through multiple disjoint shortest paths.

### D. QoE-Aware SDN/NFV-based Mechanisms using MPTCP and Segment Routing

As multimedia services/applications continue to grow, SPs have adopted SDN/NFV to implement TE to improve the network efficiency, application performance, and the end-user QoE. Notable examples include Google's B4 SDN/TE [224] and Microsoft SWAN [225]. Providing high QoE for video traffic delivery is vital for service and network providers. MPTCP has emerged as the transport protocol capable of creating multiple subflows and distributing them over multiple disjoint paths in the network such that all the links are optimally loaded. A centralized SDN controller can be used to provide an efficient mechanism for routing the MPTCP subflows in SDN/NFV-based networks. MPTCP-based SDN/NFV implementations enable load balancing, reliable communications, and better network resources utilization that lead to higher system throughput and end-users' QoE [226]. However, as stated by Duan et al. [227], the transfer of large data using MPTCP, especially multimedia traffic flows in SDN/NFV platforms, results in the number of forwarding rules on every switch, to grow tremendously. In fact, SDN switches are incapable of handling a large volume of flow rules because the complex rule matching in SDN/NFV, such as wildcards, requires switches to store rules in Ternary Content Addressable Memory (TCAM), which is expensive [228] and limited in size (e.g., practically supporting 2k-20k rules as demonstrated by [229]).

To overcome these limitations, the IETF proposed the Segment Routing (SR) paradigm to provide TE solutions by simplifying the control plane where SDN/NFV switches no longer need to maintain routing information. Storing a large number of rules is the main limitation of SDN/NFV switches which can be alleviated by using SR where a logical path of MPTCP subflows can be expressed as a sequence of segments between the ingress and egress network nodes (e.g., a switch/router/link). With SR, a host or an edge router can steer a packet through the network using an ordered list of processing/forwarding functions called segments. A segment can be a logical or a physical element such as a network node (e.g., OpenFlow switch or router), network link or a packet filter. The list of Segment IDentifiers (SIDs) forms a chain of these elements where packets are routed within an SDN/NFV system. The scope of these SIDs can be either global or local. Global SIDs are recognized by all network nodes, while local SIDs are only known to the node associated with the SID as demonstrated in [230], [231].

TCAM extensively consumes power and is usually considered as an expensive resource to be developed. Therefore, one solution to reduce the cost could be the common usage of MPTCP and SR. Such a solution would also offer flexibility, reliability, scalability and improve the end-user QoE in SDN/NFV networks. As shown in Fig 16, suppose that an MPTCP connection consists of 3 subflows: *sf1:1*, *sf1:2*, *sf1:3*, where *sf* stands for subflow and the indexes indicate the number of a subflow of a particular MPTCP connection. To guarantee the end-users' QoE-fairness level and performance requirements of SDN/NFV-based networks, the controller checks the available capacity of all connected paths and selects the shortest paths to transmit the subflows of the same MPTCP connection. As shown in Fig. 16, *sf1* will be transmitted via path, S1→S3→S2→S4, *sf2* via path S1→S7→S8→S4 and *sf3* will take path S1→S5→S6→S4.

When the MPTCP connection of subflows is established, the SDN controller can select the best available paths to transport the subflows. These paths can be expressed by the segment routing strategy into SR paths, using the segment label list as shown in [157], [232]. SDN-based MPTCP implementations can improve network throughput and meet the end-users' QoE. Indeed, [233] proposes an architecture that utilizes MPTCP in SDN-based ISP networks to provide QoS/QoE-guaranteed services to DASH clients, while an adaptive QoS/QoE differentiation based on current network feedback using MPTCP over Optical Burst Switching (OBS) in software-defined datacenters is presented by [234]. Faster download speed and improved QoE are reported by [235], where SDN is used to add or remove MPTCP paths to reduce the large number of out-of-order packets which may cause reduced performance and degrade the end-users' QoE. Wu *et al.* [236] propose a quAlity-Driven MultIpath TCP (ADMIT) approach that can achieve high quality mobile-based video streaming services with MPCTP in heterogeneous wireless networks. ADMIT incorporates the quality driven Forward Error Correction (FEC) coding and rate allocation to mitigate end-to-end video streaming distortion. Corbillon *et al.* [237] introduce a cross-layer scheduler that address MPTCP issues such as head-of-line blocking [238] during multimedia streaming services. The proposed scheduler leverages information from both application and transport layers to re-order the transmission of data and prioritize the most significant parts of the video. The viewers' QoE is maximized by decoding the video data even in difficult streaming conditions, for example when there is a small buffer and/or the video bit-rate is approximately close to the available bandwidth in the network. Ham *et al.* [239] propose a MP-DASH, a novel multipath strategy that can enhance MPTCP to support adaptive video streaming under user-specified interface preferences. MP-DASH consists of two main components, namely the MP-DASH scheduler and the video adapter. The scheduler receives video segments and the user's preferences to determine the best fetching strategy of the video segments over multipath. The video adapter is a lightweight component that handles the interaction between the DASH rate adaptation algorithms and the MP-DASH scheduler. The performance results of MP-DASH indicate that, the cellular usage and the radio energy consumption can be reduced by 99% and 85% respectively with a negligible QoE degradation comparing to the native MPTCP approach.

*E. Collaborative Service QoE Management by OTTP and ISP through SDN and NFV*

SDN and NFV may provide an opportunity to ISP and OTTP for collaborative service management. In collaborative QoE management, the OTTP and ISP have different roles. The studies conducted in [240]–[243] propose that the QoE monitoring should be performed by the OTTP because they have their applications installed in the user terminal. Moreover, the passive monitoring at the user terminal may also provide a solution to capture the context and human influence factors (not commonly available to ISPs due to end-to-end encryption) for more accurate prediction of the QoE measurements [240], [241]. The monitored KQIs of the active session is stored in the database (accessible to the ISP through RESTful API) by the OTTP. Similary, the study conducted in [244] proposes 5G network architecture for the network and service management of the multimedia services by passive monitoring probe at the user terminal for exchange of information between OTTP and MNO.

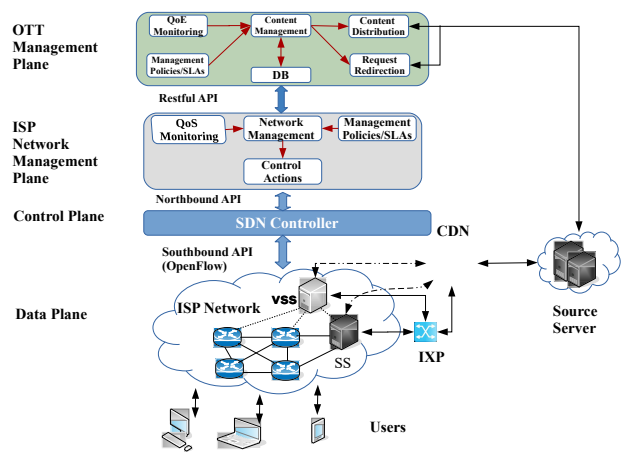

Fig. 17: Collaborative service QoE management by OTTP and ISP through SDN and NFV.

The ISP may utilize the SDN and NFV paradigm for the network management operations to drive collaborative QoE management using QoE related information provided by the OTTP. ISPs may perform network-wide QoS monitoring and network policy implementation utilizing the SDN controller. The network-wide operations such as Virtual Surrogate Servers (VSSs) placement for video streaming services, traffic rerouting, link-aggregation, and traffic prioritization can be performed in a QoE-aware fashion for the OTTP services through collaborative QoE management [242]. For example, in the case of a flash crowd appearance, VNF with VSS can be utilized to decrease the network load and end-to-end delay for content retrieval. Fig. 17 illustrates the SDN and NFV enabled collaborative service management architecture managed by the OTTP and the ISP [2], [112], [4]. The following provides the details of the architecture:

- *Data Plane*: This plane consists of the data forwarding physical/virtual network devices such as routers, switches, and VSS, which can be controlled by the SDN controller and management plane.
- *Control Plane*: The control plane is composed of the SDN controller through which the network management plane implements network-wide policies and communicates with the data plane.
- *ISP Network Management Plane*: This plane is responsible for performing network-wide QoE management and is owned by the ISPs. This plane consists of the following modules: *a) QoS monitoring*– performs QoS monitoring to monitor the network performance; *b) management policies/SLA*– considers service management policies and provides feedback to the network management module to implement network-wide policies; *c) network management*– performs network-wide QoE management by performing QoE/QoS measurements and incorporation of the management policies and; *d) control action*– implements the network policies and QoE management

TABLE V: A summary of QoE-Centric management strategies of multimedia streaming services using SDN/NFV technologies.

| Strategy | Major contributions | Contribution/Objectives/Functionality |
|---|---|---|
| Server and Network-Based Optimization | Bandwidth reservation [107], [196], [200] or bitrate guidance [7], [197], [108], [245] video stream prioritization [246], CDN orchestration and caching optimization [247], server-assisted delivery [110] | Maximizes the end users' QoE through the best path calculations for each service flow, optimal bitrates computation for the clients so as to obtain QoE-fairness in the delivered video, Optimizing the QoE by assigning a different priority to a video stream depending on the requested quality |
| QoE-Fairness and Personalized QoE-centric control | [193], [7], [205], [198], [106], [206], [207], [199], [19] | Efficient network resources allocation to ensure that the maximum users' QoE fairness is achieved. The basic idea is to avoid quality instability, unfair bandwidth sharing and network resource underutilization among DASH competing clients sharing the same bottleneck network link. |
| QoE-centric Re-routing Mechanisms | [209], [208], [213], [215], [248], [249] | Maximizes the end users' QoE through the best path calculations for each service flow. That way, the video is streamed over links with low-latency, high-throughput which increases the achieved video quality. |
| QoE-aware Cross-layer Optimization | [73], [250]–[252] | To utilize network resources efficiently and optimize the end-users' QoE through a joint cooperation between layers and coordination of their actions. In such a design, the QoE requirements can be specified at the application layer and controlled at the network layer using SDN controllers [21], [253] |
| Transport Level Optimizations | TCP/MPTCP solutions for HAS [157], [232], [233], [235] | To facilitate efficient multipath-routing and speed up the transfer of large amount of multimedia applications between end-points while guaranteeing the end users' QoE in SDN-based networks. MPTCP provides load balancing, reliable communication and better network resources utilization that leads to higher network throughput and the end-users' QoE [226], [237]. |

operation via SDN controller based on input provided by the network management module.

- *OTT Management Plane*: This plane includes the OTTP owned cloud space for the QoE related information sharing. The QoE monitoring module collects and stores KQIs in the database, which is shared with the ISP. Moreover, the content distribution and replication module assist the ISP to deploy VSS to decrease the end-to-end delay in the multimedia content retrieval.

*F. QoE-aware/driven Approaches using Full Adoption of SDN and NFV*

The integration of SDN and NFV provides means for implementing advanced network QoE monitoring and management solutions. The use of SDN and NFV provides a higher degree of freedom regarding the placement of QoE measurement points and flexible control of multimedia traffic flows compared to approaches implemented using SDN or NFV only. For example, several papers are proposing the integration of SDN and NFV in the future mobile network, and notable examples include SoftRAN [254], CellSDN [255] and the QoE-softwarized architecture presented in [125]. This section provides QoE-aware/driven approach using *full adoption* of both SDN and NFV. We define *full adoption* as QoE management solutions that integrate SDN and NFV in their implementations.

As the adoption of SDN and NFV matures towards future 5G networks, the software-defined NFV architecture [256] can offer an active traffic steering and joint optimization of network functions that further turns into QoE-management offerings [257]. Besides, use-cases that leverage SDN and NFV to provide bandwidth allocation dynamically, and QoE adaptation of video applications in home networks of an end-user are available [258]. By considering the scalability and flexibility needed in 5G systems, monitoring and discovery framework that uses the design principles of SDN and NFV is introduced in [259]. As envisaged by the SELFNET project, the corrective and preventive actions to avoid existing or potential network problems that negatively affect the end-users' QoE can be performed by SDN/NFV sensors/actuators that monitor the network. Using machine learning on top of SDN and NFV, in [257], a network resource allocation system that provides QoE-aware delivery of media services and autonomous network management to meet the changing traffic demand is proposed. A QoE-oriented self-optimization and self-healing mechanism of multimedia networks in 5G systems is proposed by Neves *et al.* [18], with much emphasis given on video quality management in the context of SDN and NFV paradigms. In an attempt to support a variety of services and the corresponding QoS/QoE requirements over SDN-NFV infrastructure, modular architecture for multi-service and context-aware adaptation of network functions is presented in [20]. The goal is to allow multiple tenants to share network resources, support on-demand allocation of radio and core resources through flexible connectivity as well as QoS/QoE management for 5G networks.

*G. Summary*

Table V summarizes the QoE-centric management strategies of multimedia streaming services. We also provide a comparison of these approaches in terms QoE-fairness on clients, deployment complexity and the most impacting metrics on QoE in Table VI. The discussion above clearly points toward a shift towards the adoption of SDN and NFV paradigms for QoE management and orchestration of network resources in future networks. It also indicates the significance of these advanced technologies for improving the performance and the delivery of QoE-rich services to the end-users in future communication systems. The inclusion of SDN and/or NFV and their significant support in future networks such as 5G is an emerging area of active research which still has a vast scope for future research.

TABLE VI: COMPARISON OF QOE MANAGEMENT APPROACHES IN TERMS OF QOE-FAIRNESS ON CLIENTS, DEPLOYMENT COMPLEXITY AND RELATED METRICS WITH MOST IMPACT ON QOE.

| Reference | Strategy | QoE-related Metrics Used | QoE-Fairness on Clients | Deployment Complexity |
|---|---|---|---|---|
| [107], [196], [200], [7], [197], [108], [246], [110], [247] | Server and Network-Based Optimization | Video chunk quality, resolution and guided bitrates, switching frequency, number of stalls, average video quality [7], [107], [110], [197]–[199]; link utilization, video instability index, startup delay, number of stall duration [196], [196], [200]; cache hit-rate, video start failure-rate, and network utilization [108] | Yes | High/Medium |
| [193], [7], [198], [106], [205]–[207], [199], [19] | QoE-Fairness and Personalized QoE-centric Control | Cost efficiency fairness, video resolution, guided bitrates, video chunk quality, switching frequency, number of stalls, startup delay, average video quality | Yes | High/Medium |
| [73], [250]–[252] | QoE-aware Cross-layer Optimization | buffer level, video playback continuity, fairness-index and stability index[260]; utility-QoE and system throughput for software-defined vehicle networks [251]; acceptance ratio, revenue-to-cost ratio and bandwidth [252] | No | High/Medium |
| [192], [260]–[264] | Application-level Optimization | video playback bitrate, segment duration, average download bitrate and buffer level [260], [263]; average video quality, buffer starvation/filling and initial startup delay, streaming encoding rate [261], [262] | No | High/Medium |
| [157], [232], [233], [235], [237], [226], [265] | Transport Level Optimization using TCP/MPTCP | PSNR, e2e delay, and goodput [226], [237]; quality switches, startup delay [265] | No | High/Medium |
| [121], [266], [267], [268], [10], [269], [60], [270], [35] [271]–[283] | QoE-Optimization using other Emerging Network Architectures | Delay, backhaul traffic load, hit ratio of cache resources [267]; average throughput, RRT latency [276], [268], buffer size [284], user density [10]; switching frequency, initial buffer delay, average video bitrates and video quality [285], [286], [280]–[282] | Only in [285] | High/Medium |
| [31], [287]–[298] | QoE Considerations in New Domains: Immersive AR/VR, Mulsemedia and Gaming Applications | Video resolution and network budget [287], screen size [289]; viewport quality and the sensitivity to head movements [299] | No | High/Medium |

## VI. QoE- Management approaches using Emerging Architectures

SDN and NFV have proved to be the appealing technologies for QoE management and control in future networks, as discussed previously in Section V. However, the introduction of emerging architectures such as MEC [14], fog/cloud computing [300], [301] and ICN can greatly benefit from SDN and NFV to provide interactive and immersive video services to consumers. Recent attempts that analyze the QoS/QoE of multimedia services offered by edge cloud include [60], [269], [270], [276] and [283], [302]. We present in this section QoE-aware/driven approaches for adaptive streaming in emerging architectures focusing on MEC, fog/cloud and ICN in Section VI-A, Section VI-B and Section VI-C respectively.

MEC promises to offer an environment characterized by high bandwidth and low latency for applications and content providers. A QoE-oriented MEC architecture enabled by SDN and NFV to support services dynamically is proposed by Peng *et al.* [35]. An efficient QoE monitoring probe that is aware of the RAN type, cell topology and resource allocation for adapting to the service delivery characteristics and end users' QoE demand is proposed in [60]. A QoE analysis of NFV-based MEC video application for the future network is presented by [269]. Authors in [302] propose a mobile edge virtualization approach with context-aware adaptive video prefetching (MVP) to achieve QoE-assured 4K video on demand delivery to the end-users. The proposed solution can enable content providers to embed their content intelligence as the VNF into the mobile network operator's infrastructure edge. An illustration of the MVP system architecture is shown in Fig. 18. The MVP edge from UE's point-of-view is set to handle video segment requests during VoD sessions. Note that the MVP edge can infer each UE and its video streaming sessions' QoE influencing factors such as video buffer status and its history per-segment download throughput. This is achieved without any feedback signaling from the UE. The MVP edge is also aware of real-time RAN context information, which is periodically disseminated through the MNO-owned MEC server's Radio Network Information Services (RNIS) module. Upon receiving the video segment request, the video segment, if available there, is served by the MVP edge with minimum access latency otherwise the MVP edge forwards the request to the original video source. The requested video segment is then retrieved and served to the UE. It is worth mentioning that besides handling the UE's request, the MVP edge also performs the following functions: (a) adaptive prefetching for each VoD session using its embedded content intelligence where video segments are downloaded in advance (e.g., during the UE's request progress), (b) traffic monitoring and prediction (c) video quality adaptation using the real-time knowledge of UEs and network. A critical aspect of this approach is that content providers can deploy their content and intelligence such as caching and prefetching policies at the MVP edge using the virtualized resources (e.g., storage, computing).

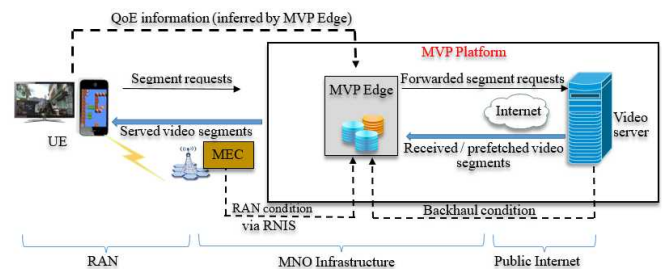

Fig. 18: Mobile edge virtualization with adaptive prefetching (Adapted from [302]).

### A. QoE-aware/driven Adaptive Streaming over Multi-Access Edge Computing

Liang *et al.* [270] present QoE-aware wireless edge caching with bandwidth provisioning and caching strategies in Software-Defined Wireless Networks (SDWNs) to decrease the content delivery latency and improve the utilization of the network resources. Li *et al.* [283] propose a MEC-assisted video delivery architecture for heterogeneous wireless networks that take into account the number of active user equipment (UE) and the available bandwidth in the network to maximize the QoE-fairness across all clients. Authors in [303] present a QoE driven mobile-edge caching placement optimization strategy for dynamic adaptive video streaming. This approach can reduce video distortion significantly for all users while taking into account the imposed constraints on the edge servers' storage capacity, the backhaul link, and the users' transmission and initial startup delay. Zhang *et al.* [271] apply the edge-computing based WiFi offloading technique to propose a QoS/QoE based resource allocation scheme over 5G mobile networks. The proposed resource allocation scheme can satisfy the joint heterogeneous statistical QoS/QoE requirements of mobile devices through WiFi offloading strategy that maximizes the aggregate effective capacity. Moreover, an Edge-based Transient Holding of Live Segment (ETHLE) strategy is proposed by [271] to achieve seamless 4K live streaming experience by eliminating buffering and substantially reducing initial startup delay and live stream latency. A demonstration of real-time QoE estimation of DASH-based mobile video applications through edge computing is highlighted by Ge *et al.* [272] while [273] presents a MEC approach to improve the QoE of video delivery service in urban spaces.

A QoE-aware Control plane for adaptive Streaming Service (QCSS) over MEC infrastructures that can assure high QoE delivery of online streaming service to mobile users is proposed in [275]. Fig. 19 shows an overview of three components of QCSS. The *client-side action component* sends the services requests to the MEC nodes and performs contextual information [9] measurements based on the collected feedback. The *edge node-side action component* performs network throughput prediction and QoE-aware bitrate adaption mechanisms in order to maximize the end users' QoE. The *center node-side action component* is responsible for selecting the most suitable edge node with the lowest cost to balance the network load and improve transfer efficiency. Moving computing and transmission loading from the center node prevent end-user clients' to access video contents from the center node simultaneously, which may cause bandwidth bottlenecks. Fig. 20 shows an active component in the client-side that is responsible for actively measuring contextual information (e.g., a set of QoS attributes, play status and activity) [275].

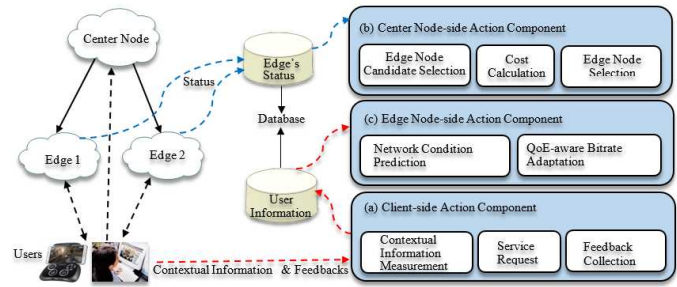

Fig. 19: An overview of QoE-aware control plane for adaptive streaming service in MEC.

### B. QoE-aware Adaptive Streaming over Cloud/Fog Computing

Fog/cloud computing [300], [301] promises a distributed computing paradigm that extends the services provided by the cloud to the edge of the network. As an extension of the cloud computing paradigm, Fog enables computing at the edge of the network (e.g., closer to IoT and/or the end-user devices). Both Fog computing and MEC support virtualization functions. However, unlike cloudlet [308] and MEC, fog computing represents the overall architecture of distributing resources across the networks and cannot operate in a standalone mode as it is tightly linked to the existence of a cloud. In the context of QoE management, Gupta *et al.* [274] propose a software-defined FC (SDFog) architecture that can perform QoS/QoE-aware orchestration of resources by scheduling flows between services. SDFog is a service-oriented approach for enabling resource management with end-to-end QoS/QoE requirements in a heterogeneous FC environment. SDFog consists of two major components namely, the Service Oriented Middleware (SOM) and the Distributed Service Orchestration Engine (DSOE). The DSOE is responsible for service discovery, flow creation and network parameter calculations, and QoS aware orchestration of resources by scheduling flows between the services.

Based on SDN and NFV, architectural proposals for 5G services to perform orchestration of multi-technology and multi-domain networks on top of a cloud/fog infrastructure are demonstrated in [304] and [305]. In the context of 5G networks, a unified Network Service Chain (NSC) strategy in SDN and NFV that can perform fast computation of services on the cloud computing and FC model is presented by [306]. Rosário *et al.* propose a SDN-based multi-tier fog computing architecture that provides the cooperation between fog and cloud to run video services with QoE support. In order to minimize the traffic in the core network, authors propose to move video services from cloud computing to fog nodes. The major benefit of this approach is that it ensures that mobile users have access to fog services with low delay, QoE support, and without significant network overhead.

---

[9]contextual information is the set of QoS attributes, play status, locations and activities which are relevant to support the streaming service [275].

TABLE VII: A SUMMARY OF QOE-CENTRIC MANAGEMENT STRATEGIES IN EMERGING ARCHITECTURES.

| Strategy | Major contributions | Contribution/Objectives/Functionality |
|---|---|---|
| Application-Level Optimizations | Adaptive streaming over HTTP/2 [192], [260]–[262], client-based prefetching [263], Meta-heuristics for increased client QoE-awareness [264], [60] | Improving the video quality and reduce the live latency using HTTP/2 protocol. HTTP/2 is a server push mechanism that also increase link utilization compared to HTTP/1.1 [260]. Meta-heuristics approaches exploit context information to improve the bitrate selection strategy of the client. |
| QoE-aware/driven Adaptive Streaming over MEC | QoE-aware software driven multi-access edge service management [35], [266]–[268], [271], [273], [275], [303] | Decrease the content delivery latency and improve the utilization of the network resources. Enable QoE monitoring and video quality adaptation using the real-time knowledge of UEs and network. |
| QoE-aware Adaptive Streaming over Cloud/Fog computing | QoE-based resource management [274], [276], [277], [304]–[306], QoE optimization with energy efficiency [285] | Perform QoS/QoE-aware orchestration of resources by scheduling flows between services. Some of the proposals such as in [284] can enable service providers to predict the QoE of DASH-supported video streaming using fog nodes. |
| QoE-driven/aware Management using ICN | [279], [281], QoE-driven content caching and adaptation scheme over MEC-enabled ICN [303] | Performing prefetching of video streams that enables ICN to compute the link resources availability and makes scheduling of data units dissemination called "chunks" to edge caches according to end-users' requests [307] or video prefetching at the network edge in order to achieve the users' QoE [302] |

Kitanov and Janevski [276] present a QoE evaluation of Fog and cloud computing service orchestration mechanisms in future networks. Important parameters considered for QoE evaluation include throughput, latency, and energy efficiency per user for different payloads. While cooperation fairness is a critical issue in fog computing, [285] presents a joint optimization of energy and QoE with fairness in cooperative fog computing. A QoE-aware application placement policy that calculates the capabilities of Fog instances considering their current status and prioritizes different application placement requests according to user expectations is presented in [277]. MEdia FOg Resource Estimation (MeFoRE) is one of the QoE-based framework proposed by [278] to provide resource estimation at Fog and to enhance QoS in IoT environments. Zheng *et al.* [284] propose a Fog-assisted Real-time QoE Prediction (FRQP) strategy to enable service providers to predict the QoE of DASH-supported video streaming using fog nodes. FRQP uses the network-measured traffic to observe packet header information of the video traffic. A probe is deployed at fog nodes to infer users' QoE according to the temporal features of the video traffic flows. For a more recent survey of delay-sensitive video applications (e.g., video conferencing) in the cloud, we refer the reader to [309].

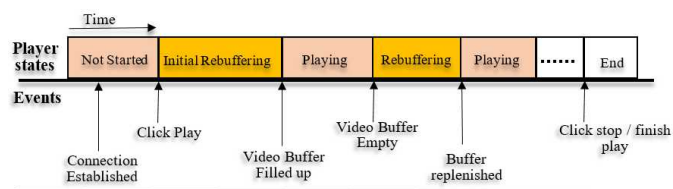

Fig. 20: Play status and activities of streaming service in client side (adapted from [275]).

### C. QoE-driven/aware Management Approaches using Information-Centric Networking (ICN)

ICN [310], [311] is a disruptive network architecture that proposes to shift from the traditional host-centric approach model to a content-centric model by directly naming and operating on information objects [312]. In ICN, contents are decoupled from the location at the network level such that when a client requests a specific content, it is identified by a Uniform Resource Identifier (URI) using a routable name scheme, for example, /example.com/video/ICNVideo.mp4. The clients have full control over the streaming session and also the possibility to adapt the multimedia stream to its context (e.g. device capabilities, network conditions) [313]. Data-Oriented Network Architecture (DONA) [314], Content-Centric Networking (CCN) or Named Data Networking (NDN) [315] are typical instantiations of the ICN paradigm [310].

Yu *et al.* [281] propose a DASH-aware video stream prefetching approach in ICN and CDNs where the network controller is used to locate content, manage the cache, and monitor network congestion. Based on the collected information regarding the usage of network capacity and DASH video streaming sessions, the network controller performs prefetching of video segments and move them to the edge caches. Li *et al.* [279] use a binary integer programming formulation to present dynamic adaptive streaming over popularity-driven caching (DASCache), an approach that can handle different video content caching in ICNs. DASCache enables fast video streaming with high QoE over ICNs by minimizing the average access time per bit of video contents requested by the end-user. DASCache can also enable users to switch to videos with better resolution. This allows them to achieve the best video watching experience.

Authors show that using DASCache, the best experience for users is achieved in varying network conditions by switching to videos with better resolution. Focusing on DASH-based video applications, Ge *et al.* [316] propose a QoE-driven content caching and adaptation scheme over MEC-enabled ICN-based architecture. The main idea is to handle the popularities of both video segments and their representations at the mobile network edge that ultimately enhance the users' QoE. Su *et al.* [317] propose a game theory-based layered videos approach in mobile social networks that can optimize resource allocation from cache nodes and enhance the video delivery performance for mobile users. The proposed scheme can achieve a higher hit ratio and users' QoE for multiple social groups with limited capacity. Authors in [303] present a QoE-driven mobile edge caching placement optimization strategy to reduce total video distortion by taking into account the coordination among distributed edge servers and rate-distortion characteristics of multiple bitrate videos.

As shown by Lederer *et al.* [280], the video rate adaptation of HAS clients can also be complicated by ICN. This is so because the client in ICN is not aware of the node, such as the original server that provides the content. Liu *et al.* [318] propose DASC, a dynamic adaptive streaming system that implements DASH over CCN. Authors report that the DASC overcomes the network bottlenecks by converging progressively to the best available video quality. The interplay between different adaptation heuristics and interest forwarding strategies in NDN [282] are demonstrated where an interest represents the client request for specific content. The upper bound for the average bitrate the clients can obtain is found, assuming that the network and streaming characteristics are known a priori. With similar objectives as regular DASH, it is possible to develop intelligent caching mechanisms in ICN [281], [319].

It is worthwhile mentioning that caching in ICN use the naming structure of the video packet interests. That way, the network knows what a client is watching, and also the relationship between different video segments [281]. Yu *et al.* estimate future segment requests to prefetch a video content during off-peak congestion periods by using network condition information available at the ICN nodes. Using video- and network-aware prefetching strategy, the delivered video quality increases by 20%, compared to a DASH system without prefetching. Moreover, video quality improvements can be obtained when considering that each network node has caching functionalities in ICN. In this aspect, a video client can opportunistically retrieve video segments from both the server, using 3G/4G, and from other clients in a peer-to-peer fashion, using Wi-Fi. This solution results in an increase in quality and reduced load on the mobile network.

### D. Summary

Table VII provides a summary of QoE management strategies of multimedia services using emerging architectures (MEC, ICN and cloud/fog computing). Multi-access edge decrease the content delivery latency and improve the utilization of the network resources. Cloud computing provides ubiquitous on-demand access to a shared pool of configurable computing resources (e.g., networks, servers, storage) with minimum management effort. ICN can improve the end users' QoE through edge cache prefetching mechanisms that consider network monitoring for populating a cache as per the requirements/needs ahead of the client requests. It is worth noting that ICN approaches may also combine the benefits of pervasive fine-grained caching platforms and multi-path transmissions similar to MPTCP, an advantage that entails higher throughput compared to standard TCP/IP.

## VII. QoE in Emerging Applications: Immersive AR/VR, Cloud Gaming, Light Field Display and Mulsemedia

This section provides QoE studies in emerging immersive applications focusing on augmented reality (AR) and virtual reality (VR), MULtiple SEnsorial MEDIA (MULSEMEDIA), cloud gaming and light field applications.

### A. QoE in Immersive AR/VR and Mulsemedia Applications

With the increasing acceptance and demand of 360° immersive videos which are of much higher size than traditional 2D videos, Internet traffic is expected to further grow in the next few years. For example, the global market of 360° cameras is expected to grow at an annual rate of 35% between 2016 and 2020[10] while the global market of VR related devices will reach 30 billion USD by 2020[11]. As new use cases for AR and VR are introduced in the market, service providers will need to address bandwidth limitations, reduce end-to-end network latency, and improve the overall QoS/QoE for the streaming media services. These challenges not only impact the consumers' AR/VR experience today but remain an impediment to the delivery of more immersive experiences in real-time in the future.

With the expected increase of personal HMDs and new content generation devices, researchers have recently started to quantify, model, and manage QoE when the user consumed content is beyond traditional audio and video materials. In these approaches, viewport-dependent solutions [287], [288] have been proposed for 360° video streaming, because they can reduce the bandwidth required to stream the video. The user in VR is immersed in a virtual environment and can dynamically and freely decide the preferred viewpoint [31]. In viewpoint-dependent streaming, the portion outside the viewpoint is streamed at an average or lower quality while the part of the video watched by the user is streamed at the highest possible quality. It is worth mentioning that viewpoint-dependent streaming can be obtained using online transcoding operations, such as using foveate-based encoding [287], or by spatially tiling the video [288].

QoE measurements has also been addressed in the context of AR systems [289], [290] because of their applications in various fields such as emergency response training and medical surgery. Gandy *et al.* [321] use a physiological approach to study QoE measurements where a three-lead electrocardiogram (ECG) sensor was placed on the subjects' chest. Authors use also a skin temperature sensor and galvanic skin response (GSR) mounted on subjects' non-dominant hand. The results from this study indicate that the frame rate have a smaller effect in VR applications with regards to improving the end-user's QoE. Perritaz *et al.* in [322] propose a video bitrate adaptation approach to maximize the end-user's QoE by adapting the frame rate and the image size in real time. QoE aspects (e.g., subjective quality assessment, and psychophysics, usability, human factors, ergonomics, ethnography) in AR applications are presented by Puig *et al.* in [289]. Pallot *et al.* [290] investigate a collective user experience when multiple users watch the same sports game via AR technology. Pallot *et al.* further proposes a taxonomy of QoE in augmented sports and a 3D LIVE project using QoS-UX-QoE approach that can be applied in three use cases (skiing, jogging and golfing). A discussion of QoE issues in the implementation of AR services in big cities with a high density of users is given in [291]. The impact of stalling events in a fully immersive setting involving users watching omnidirectional videos using an HMD is discussed by Schatz *et al.* in [323].

---

[10]https://goo.gl/zJCdnO
[11] https://goo.gl/nw9mtP

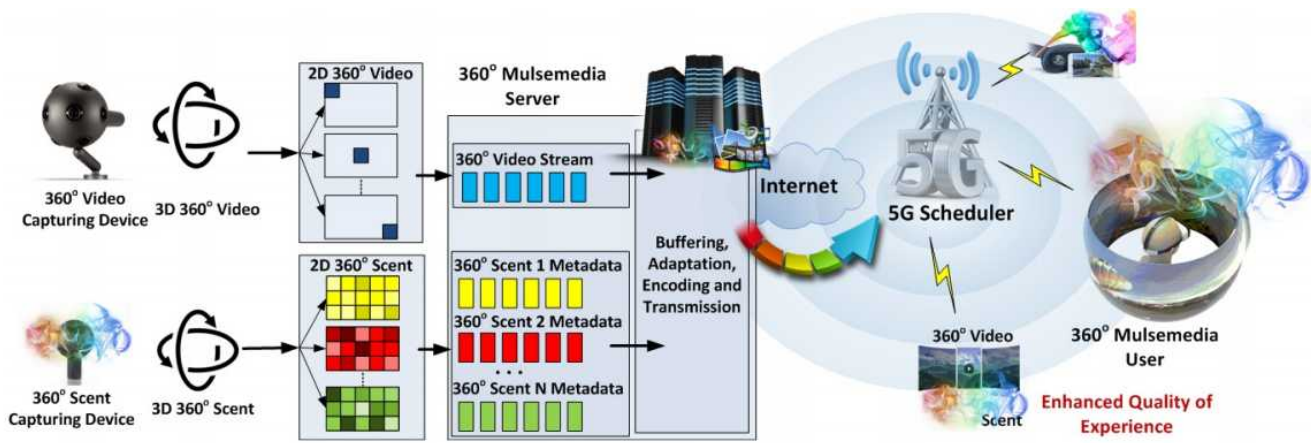

Fig. 21: 360° mulsemedia experience delivery system over next generation wireless networks [320].

QoE in emerging mulsemedia services, which involve different new media objects than traditional multimedia applications, has recently attracted much attention from the academia and industry [292], [293]. Mulsemedia enables the integration of other human senses (e.g., olfactory and haptic) into human-computer interaction. Thus, 360° video applications, for example, could integrate additional 360° sensory media content (e.g., olfactory, haptic or thermoceptics media objects) that could eventually enable an even better quality of user experience [324]. However, the new 360° mulsemedia services would come at the cost of more bandwidth than the conventional applications [298] and stringent delay requirements [299]. Yuan *et al.* introduce QoE of mulsemedia services where users can inform the mulsemedia server about both user preferences and network delivery conditions [294].

The impact of intensity of certain mulsemedia components, including haptic and air-flow on user-perceived experience are extensively studied in [295], where making use of mulsemedia increases the overall user enjoyment levels by up to 77%. Jalal *et al.* [296] propose a nonlinear model for predicting the QoE of high dynamic spatio-temporal mulsemedia. The authors focus on the QoE of audiovisual sequences enriched with additional sensory effects such as light, wind, vibration, and scent. QoE beyond audiovisual referred to as sensory experience is introduced by Murray *et al.* in [297] where sensory effects such as ambient light, scent, wind, or vibration are utilized as additional dimensions that contribute to the end users' QoE. A dataset of videos enriched with olfactory content and annotated with subjective user ratings are also provided. A new concept of 360° mulsemedia as the application that will revolutionize the streaming technology is introduced by Comsa *et al.* [320]. Fig. 21 shows the proposed 360° mulsemedia delivery system. At the server side, alongside the 360° video capturing device, a 360° scent capturing device is also able to collect various olfactory types associated with the video representation. The 360° mulsemedia server has several functionalities such as 360° olfactory objects mapping, 360° media objects synchronization, buffering, adaptation encoding and transmission. The 360° mulsemedia content is transmitted to the 360° mulsemedia user over the 5G wireless networks. The performance of the 5G radio scheduler is shown to depend on the number of 360° mulsemedia users, mobility, positioning and channel conditions.

To further enhance the appeal of VR services, 360° VR video streaming contents that deliver a more immersive experience is being developed. The 360° video streaming and immersive video streaming will enable mobile operators to attract users by providing competitive service contents during major events (e.g., sports games, artistic performances, etc.). It is worth mentioning that, 360° live VR will enable users to participate in the entertainment events without going anywhere. While there have been significant research efforts regarding QoE assessments in AR/VR [287], [288], the VR-oriented E2E network operation and management system become critical for understanding the user perceived quality of immersive multimedia experiences. This is so because the overall process of delivering the VR video content from the source to clients is more complex compared to that of common 4K videos. The VR-video quality can be degraded by various faults which can be hard to separate or distinguish from. This necessitates the development of effective QoE-oriented E2E solutions at various points of VR-video delivery system for real-time monitoring, detecting and demarcating faults. This would enable service providers to enhance VR-video streaming service experience through a multi-sensing VR QoE solution that can model user experience and measure the perceivable media quality during future softwarized network transmission [325].

### B. QoE in Cloud Gaming Video Streaming Applications

Gaming video streaming is becoming popular with the rise of two related and accessible services. One such form of service is passive gaming video streaming where the gameplay of players is streamed from the client to streaming sites such as Twitch.tv, YouTube-Gaming and Facebook Gaming. The second form is the cloud gaming service where the game is rendered on the cloud, and the gameplay of users are streamed instantly to the client. One example of such a cloud gaming service is GeforceNow by Nvidia. Both services have different network requirements and constraints, which require different network management schemes. Cloud gaming is a delay sensitive multimedia application. Passive gaming video streaming suffers from characteristics of the traditional live/on-demand video streaming services.

Cloud gaming can benefit a lot from new emerging technologies such as SDN by providing an optimized flow distribution. Some researchers have been focusing on designing the SDN controller to provide lower latency, and hence consequently, higher QoE. Amiri *et al.* in [326] proposed an optimization model that takes into account the game type, server load, and delay of the current path in order to optimize the flow distribution that minimizes the delay within the cloud gaming data center. In order to reduce the complexity of the optimization method, authors in [327] proposed a Lagrangian Relaxation heuristic method, which then can be implemented in the data center using the OpenFlow controller [327]. Although SDN can provide benefits to delay-sensitive applications such as cloud gaming, there is a need to analyze the potential tools and techniques which can help in further reduction of overall latency.

In order to standardize such emerging services, there are a few ongoing standardization works concerning the subjective and objective quality assessment of video gaming streaming services. Towards this end, two ITU-T recommendations are published [328]. One covers the identification of factors affecting QoE in gaming applications (G.1032) [329] while the other describes the definition of subjective methods for evaluating the quality experienced during gaming activities (P.809) [330]. Also, within ITU-T Study Group 12 another work item, G.OMG was established with the aim of developing a QoE-based gaming model for predicting the overall quality based on the characteristics of the network, system, as well as player and usage context factors. In addition, there are several research works regarding objective and subjective quality assessment of gaming video content such as creation of gaming video datasets [331], evaluation of existing metrics [332]–[335] and development of new no-reference metrics and models [46], [336], [337] for gaming content.

### C. QoE in Light Field Applications

A light field is a vector function which can be described as a 5D plenoptic function describing the space of all possible light rays, with radiance representing the magnitude of each ray. A temporal sequential collection of light field images can be considered as a light field video file. As compared to traditional Broadcast or streamed video content, light field videos usually have very high resolution (typically 50-80 megapixels) and hence have very high bandwidth requirement which calls for more efficient compression as well as higher network bandwidth. Additionally, light field 3D displays require different measurement procedures compared to the conventional stereoscopic displays as it includes the measurement of certain additional parameters. This is because a hologram-like image is produced by using geometrical optical techniques. Wang *et al.* [338] perform QoE measurement for light field 3D displays on Multi-layer light field displays (MLLFDs). To measure accurately the spatial resolution, viewing angle, and depth resolution, authors further propose three customized virtual models, namely the USAF-E model, the view angle model, and the concave/convex object model. A new full-reference quality assessment model for stereoscopic images is proposed by Shao *et al.* [339], where binocular receptive field properties are learned and aligned with human visual perception. Chen *et al.* [340] measured the binocular quality perception in the context of blurriness and blockiness. The global luminance similarity (GLS) index is computed by considering the luminance changes. The Sparse Feature Similarity (SFS) index is calculated by considering the amplitude difference phase and amplitude difference of sparse coefficient vectors [340]. Perra [341] presents the QoE evaluation of light field applications when viewing rendered decompressed images. A metric for quality evaluation of the rendered views that measures the variation of structural similarity on a set of viewpoints extracted from the light field is proposed [341].

### D. Summary

The industry and academia has recognized and accepted virtual reality and augmented reality as future applications that will provide a truly immersive and interactive multimedia experiences to the end-users. The recent advancements regarding AR HMDs such as mobile ready Samsung Gear VR, Epson Moverio BT-300, PlayStation VR and Microsoft HoloLens can allow users to experience their real-world environments. The discussion presented in this section indicates that, there are many interests from the multimedia community to investigate the QoE aspects to new domains such as immersive augmented and virtual reality, mulsemedia, video gaming and light field applications. Most of the presented approaches investigate the QoE evaluation for users watching videos while using a AR/VR device. Some of the works consider the subjective quality assessment and psychophysics, usability, human factors, ergonomics, ethnography, etc. in AR applications. The evolving concept of 360° mulsemedia as the application that will revolutionize the streaming technology in future 5G networks is also presented. Another domain of gaming video streaming which has gained much acceptance from the users is also discussed and presented. Majority of works in this new domain has been towards investigating the subjective and objective quality assessment of video gaming streaming services and development of high performance no-reference quality metrics. QoE measurement for light field applications considering 3D displays on multilayer LCDs or binocular receptive field properties which are learned and aligned with human visual perception is also presented.

## VIII. QoE Management Challenges and Research Directions in Future Softwarized Networks

The maturity and the inherent potentials of SDN, NFV and other emerging technologies (MEC and cloud computing) are paving the way for transforming the future network infrastructure. These technologies promise to offer a flexible network deployment and operational improvement as well as facilitating optimal use of resources in the network for QoE provided to the end-users as envisaged by NFV industries [152]. However, several new research challenges are emerging which the multimedia management community needs to address as SDN and NFV mature. This section provides a discussion of the significant future challenges and the newly arising research opportunities that can be investigated by those focusing on the QoE-aware adaptive streaming and future network management domain. We present QoE management and orchestration challenges in Section VIII-A. This is followed by a description of OTT-ISP collaborative service management using SDN and NFV in Section VIII-B. Section VIII-C and VIII-D present challenges with regards to QoE-oriented business models and QoE-based big data strategies in future softwarized networks. Section VIII-E provides QoE-oriented network sharing and slicing while Section VIII-F provides a description of scalability, resilience, and QoE optimization. The challenges for multimedia communications in IoTs is given in section VIII-G.

### A. QoE Management and Orchestration Challenges

Moving legacy NFs from a hardware-centric to software-centric approach using SDN and NFV in future networks not only demands changes on how networks are deployed, operated and managed but also on the orchestration of resources while making sure that the network functions are instantiated in a systematic and on-demand basis. Towards this direction, the ETSI MANO framework has already shown a direction, with anticipated capabilities of life-cycle management and configuration of VNFs. Following that trend, other proposals have appeared that provide solutions for a management platform for VNFs such as Cloud4NFV [342], or NetFATE [343] (which considers the desired QoE of traffic flows during the orchestration of virtualized functions). Relying on the ETSI NFV framework, the AT&T's ECOMP project [344], the Open Source MANO (OSM) project [345], and the ONAP project [346] implement the Service Orchestrator (SO) on top of NFVO. ONAP provides a vendor-agnostic, policy-driven service design, implementation, analytics and life-cycle management for large-scale workloads and services, such as residential vCPE. With ONAP, operators can orchestrate both physical and virtual NFs synchronously. The OPNFV [185] creates a reference NFV platform to accelerate the transformation of enterprise and service provider networks. Still, the OSM [345] from ETSI NFV working group is working on a reference framework that implements MANO functionalities by integrating three other open source platforms (OpenMANO [347], RIFT.ware [348], and JUJU [349]) into a single platform. Other related MANO frameworks and architectures that consider the management and orchestration of both virtualized and non-virtualized functions have been proposed in [20]. Despite these efforts, current proposals are only focused on NFV management and orchestration. There are no efforts given on the management and orchestration of both SDN and NFV resources for future evolution, for example, considering 5G networks. While existing projects are focusing on novel architectures that provide the needed flexibility and programmable networks using SDN/NFV, QoE-aware/driven management schemes for multimedia delivery services over future softwarized infrastructures are not covered yet. Moreover, to improve the decision making of NFV MANO specifically using policy/intent based networking (PBN), analytics and big data approach for managing softwarized networks have to be developed.

The emergence of SDN, wireless network virtualization (WNV) and cloud radio access network (C-RAN) provide a connecting and powerful network management mechanism for future heterogeneous wireless network environments [358]. Similar to NFV concept, WNV aim to provide efficient resource utilization, reduces CAPEX and OPEX to service providers and better QoE to the end-users in future cellular systems. In Virtualized Wireless Networks (VWNs), the radio resources and physical cellular infrastructures are owned by Infrastructure providers (InP). The MVNO is responsible to create and operate leased virtual network resources from InP. The MVNO utilize the network slices and provides services to the corresponding users without knowing the fundamental physical network architecture. Rawat *et al.* [359] investigate wireless virtualization where resources are adapted based on the demands from the users. The user utilities are subjected to QoS/QoE requirements such as mobility, coverage, rate and delay. Despite the potential benefits that come with network virtualization as stipulated by [360] and [361], several research challenges about QoE management need to be resolved through a comprehensive research effort before its full deployment. Some of the challenges include QoE-aware resource discovery and isolation and QoE pricing-based allocation and security [361]. To this end, proper and more efficient QoE-aware resource management, QoE-aware scheduling [362] and the whole QoE-based cross-layer software-defined approaches should be designed and implemented.

### B. OTT-ISP Collaborative Service Management in Softwarized Networks

The collaborative service management by the OTTP and ISPs faces many challenges which are mainly composed of monitoring and management of the collaborative service depending on the role of each entity. The OTTP may monitor the QoE of its application using passive probes at the user terminal. Based on the solution proposed in [240] and [69], the passive probe can be shared with the ISP. However, the selection of the appropriate monitoring probe frequency remains a big question. The primary reason is that if the passive monitoring probe at the user terminal operates at a higher frequency, the prediction of the QoE may be more accurate [82] but will have a negative impact on the user end device regarding utilization of the device resources (e.g., CPU, RAM and battery power) [240].

TABLE VIII: A SUMMARY OF QoE MANAGEMENT CHALLENGES, CURRENT CONTRIBUTIONS AND RESEARCH OPPORTUNITIES IN FUTURE SOFTWARIZED NETWORKS.

| Challenge | Current Contributions | Research Opportunities |
|---|---|---|
| OTTP and ISP Collaboration | [2], [112], [116], [240], [242] [241], [4], [243], [244] | The collaborative QoE-aware service management solutions including the reference architecture, optimization algorithms for the service management and business models. |
| Emerging Multimedia Applications | [31], [287]–[289], [321], [292], [293], [336], [332], [333], [335], [341] | Mechanisms for ensuring the QoE for VR/AR, Mulsemedia, Video gaming and light field display. The current QoE models for delivery of adaptive video streams have three limitations: (1) they are developed to capture the behavior of the "average" user, and hence some of them are not personalized, (2) they do not consider the context in which the streaming session takes place, and (3) only the QoE model of the users is inserted into the control loop, but not the user herself [31]. Therefore, using the network, application and user-level parameters could potentially allow creating QoE personalized models of emerging multimedia services. |
| Management and Orchestration | [7], [60], [185], [107], [342]–[352], [19]–[21], [246] | The management and orchestration of both SDN and NFV resources in the context of FNs. QoE-aware/driven softwarized management schemes for multimedia delivery services in FNs are not covered yet. |
| HTTP Adaptive Streaming over MPTCP/QUIC, Immersive Video Streaming | [157], [226], [232]–[235], [237], [353], [354] | More investigation on the impact of MPTCP/QUIC and SR on adaptive streaming over softwarized 5G networks is needed. For immersive video streaming, viewport-dependent solutions for VR streaming in future communication systems have to be investigated more. |
| Video Encoding | [355]–[357] | Many current video encoding strategies are focusing on improving existing codecs or develop newer encoders to achieve higher compression efficiency especially for new contents (e.g., HDR, AR/VR and video gaming), so as to reduce the required transmission bandwidth. However, with the arrival of IoT and M2M communications in 5G networks, it is imperative for the industry and academia to come up with newer solutions catering to the changing requirements of such applications (such as low-delay, low power and low complexity encoders). |

For the QoE-oriented service management, a continuous exchange of information between OTTP and ISPs is needed which may require interfaces for the information exchange [4]. The cloud and peering exchanges may provide an opportunity for information exchange among the OTTP and ISPs. The work in [4] proposed the use of cloud databases for the storage and retrieval of information exchange between OTTP and ISP to perform control actions in case of quality degradation. However, research needs to be conducted to validate the viability of the solutions by taking into account the E2E delay involved in the data retrieval from the cloud/fog databases until the activation of the control actions. Future research work in this direction is critically needed as the higher operating frequency for the information exchange may lead to improved service resilience while causing increased network overload issues. Concerning SDN and NFV based collaborative service management, the optimization of the OPEX and CAPEX remains an open challenge. Another challenge arises regarding content distribution/replication of the OTTP content in the VSSs of the NFV based ISP architecture related to cost of operation, load balancing, replication of the content and latency of content retrieval especially in case of flash crowd appearance.

The OTTP-ISP collaborative QoE-aware service management may also require SLAs/ELAs and business agreements among the providers for the service management policies, which may lead to service provision into different *Class of Service* to different users. The collaborative QoE management of the service may also require QoE prediction models which can provide predictions for longer duration videos rather than shorter duration. Additionally, algorithms for the collaborative QoE-aware service management are also needed where multiple-domains of the service management should be considered, such as QoE-fairness and business models.

### C. QoE-oriented Business Models in Future Softwarized Network

The emerging applications in future softwarized networks require new QoE-oriented business models to be used by the service providers in the proposed contract with their customers. SLAs used today are not enough as the means to provide QoE-related contracts between service providers and customers. Web services, has become a mainstream necessities for the formalization of new QoE-oriented contracts between service providers and end-users. Unfortunately, the current SLA modeling proposals are mostly confined and focused on QoS technical aspects. Moreover, they do not follow web principles and semantic approach for QoS specification of communication services using QoE parameters.

To ensure the Quality of Business (QoBiz) [12] in future communication ecosystem, ISPs should focus towards QoE marketization through ELAs which go beyond current QoS-oriented SLAs. This will enable to foster new business practices with a minimum QoE guarantee to the end users. To manage the QoBiz in future networks effectively, a business reference model shown in Fig. 22 can be a starting point towards developing a new QoBiz model for service providers and customers. The model integrates the ELAs, as a QoE-oriented augmentation beyond the traditional SLAs which we believe will enable new business provisioning with QoE-differentiated services to the end-users. We argue that future SLAs should not be defined based on QoS as in today's networks. Instead, they should be described on ELAs [364], mechanisms which are purely based on QoE for interfaces that require ELA definitions. This will enhance and simplify the planning process of applications/services by modeling future systems' performance concerning the users' experience. While the user churn remains the most critical to all the businesses in the multimedia services, the correlation of QoE and user churn is still another area that needs investigation. An objective utility function of the user churn as a mathematical function of the predicted QoE delivered to the user is proposed in [2]. However, the subjective validation of the user's QoE churn model is still required. Furthermore, future research to investigate the impact of service pricing on the user churn, user satisfaction, willingness to pay and perceived quality is still an open question. Similarly, the study conducted in [244] proposes zero-rate QoE approach for the radio resource management in the 5G networks while considering multimedia services. This work presents zero-rated QoE as an alternative to the already commercial used scheme known as zero-rated data rate where OTTP and MNO collaborates for QoE-aware network management. The approach outperforms the traditional zero-rated data rate approach in terms of the network resource utilization, delivered QoE and QoE-aware fairness.

[12]The QoBiz quantify the business revenue from a service offered over the Internet. QoBiz parameters are expressed in monetary units such as cost of servers, revenue of website, transactions loss or dollars per transaction [363].

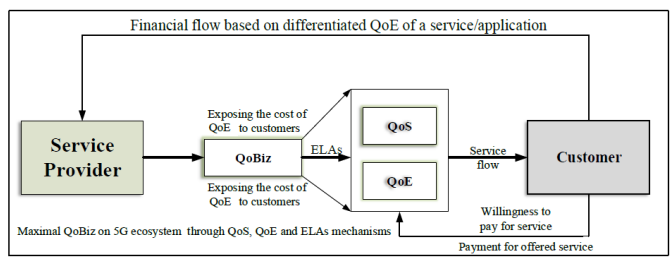

Fig. 22: QoS, QoE, ELAs and QoBiz relationship model for future communication systems.

### D. Intelligent QoE-Based Big Data Strategies in Future Softwarized Networks

As of today, a tremendous amount of data is being generated from different sources such as the IoT, social networking websites (Facebook, Twitter, and Flicker), etc. which is bound to increase even more in the coming years. It is important to note that at this pace, the current static measurements of network and application performance will not be capable of keeping up to the changing dynamic landscape of future network softwarization. Therefore, creating a QoE-based dynamic model to correlate the resulting big data, probably, requires Machine Learning (ML) and Artificial Intelligence (AI) that will move from the traditional lab-based modelling towards a QoE-driven live network predictive data analytics which is required for self-optimization and self-healing in future networks.

As stated by Cui *et al.* [365], on the one hand, SDN can solve many issues of big data applications (e.g., big data acquisition, processing, transmission and delivery in cloud data centers). On the other hand, as an essential network application, big data will have a profound impact on the overall operation and design of future SDN-based networks [366]. For example, Wang *et al.* [391] introduce a cross-layer modular structure for big data applications based on SDN. Specifically, the run-time network configuration for big data applications is studied to jointly optimize the network utilization and application performance. Monga *et al.* in [392] introduce SDN to big scientific data architectural models while an approach to big data analysis in SDN/NFV-based 5G networks is introduced by Barona *et al.* in [367]. However, while it will be challenging to meet QoS and QoE requirements and orchestrate VNFs without big data analytics, we note that the relationship between SDN, NFV, and big data is not yet studied, especially in the perspective of future networks. Therefore, it is vital to investigate new strategies for assigning and managing resources (e.g., in cloud data centers) to meet the SLAs/ELAs of various big data applications in future networks.

### E. QoE-oriented Network Sharing and Slicing in Future Softwarized Networks

Moving from hardware-based to software-based platforms could potentially simplify the multi-tenancy support where multiple services/applications from different vertical-specific use cases can be accommodated over a common SDN/NFV-based infrastructure. Although the dynamic resource sharing among slice tenants would make network resource utilization more efficient, it needs intelligent scheduling algorithms to allocate resources among these slices. Besides, the problems concerning NFs placement within a slice, intra/inter-slice QoE management still needs significant efforts to achieve and realize the effectiveness of the network slicing management in future networks [36].

Also, another research direction that needs extensive exploration is related to the isolation between slices, mobility management, dynamic slice creation, and security [384], [36]. Concerning isolation, a set of consistent policies and appropriate mechanisms have to be clearly defined at each virtualization layer. On the other hand, regarding performance, specific service performance and QoE requirements have to be met on each slice, regardless of network congestion and performance levels of other slices. With security and privacy, efficient mechanisms have to be developed to ensure that any attacks or faults occurring in one slice must not have an impact on another slice [380], [393]. That way, the network sharing and slicing in future softwarized networks using SDN and NFV can be realized in a practical implementation.

### F. Scalability, Resilience, and Optimization in Future Softwarized Networks

Many challenges in SDN such as scalability issues [132], resilience [394] and robustness of the control plane as well as reliability [395] that can negatively affect the end users'

TABLE IX: A SUMMARY OF RESEARCH DIRECTIONS AND RECOMMENDATIONS IN FUTURE SOFTWARIZED NETWORKS.

| Research Topic in SDN/NFV | Research Challenges and Recommendations |
| --- | --- |
| Strategies for Big Data [365]–[367] | Creating a QoE-based dynamic modeling procedure that leverage on big data extracted over softwarized networks. Revisit the relationship between SDN, NFV and big data in the context of future networks. |
| Scalability, Resilience and QoE Optimization [368]–[375] | Developing a QoE-aware multi-objective optimization model for both, SDN controllers and VNFs/VNFMs placement in future networks. Develop QoE-based resource allocation algorithms that consider multi-domain and distributed VNFs, dynamic QoE-based resource management and network survivability over softwarized networks. |
| Network Sharing and Slicing [20], [376]–[383] | The problems related to the placement of NFs within a slice, slice orchestration, or inter-domain services, slicing need to be further studied to achieve the effectiveness of NS over software defined/driven networks. Again, isolation between slices, mobility management, dynamic slice creation and security [384] aspects need major research efforts. |
| QoE Business Models [5], [385] | Developing new QoE business models. This is so because, SLAs used today are not appropriate as the means to provide QoE-contracts between service providers and customers in FNs. |
| Network Performance, Evaluation and Benchmarks [386]–[388] | Developing performance evaluation methodology and benchmarking tool (e.g., TRIANGLE [387] and Open5GCore [386]) that would help application developers and device manufacturers to test and benchmark new applications, devices, and services. It is important also to develop virtualisation testbeds such as OpenSDNCore [388] that will provide practical implementation of the future network evolution paradigms leveraging NFV/SDN environment. |
| Security, Privacy and Trust [180], [389], [390] | The impact of traffic encryption on users' QoE over softwarized networks is an urgent area that needs investigation. Same way, new personalized QoE models in FNs that include privacy and security together with classical video streaming metrics over SDN/NFV have to be developed. |

QoE has to be well investigated. On the controller design viewpoint, the centralized controller implementations suit a single network domain. However, the risk of becoming a Single Point of Failure (SPOF) increases, which in turn compromises the network reliability and its performance as well as the end-users' QoE. Also, the centralized controller design might create further scalability issues, especially in IoT-like networks with a large number of hosts [368]. While that is the case, the distributed controller design that logically maintains a centralized view [369], becomes the only solution to meet high reliability, scalability requirements in SDN-based networks [370]. Providing a reliable and scalable softwarized network would also translate to service provisioning to the end-users with better quality. However, the design of a physically distributed system raises practical challenges such as (1) determining the required number of SDN controllers and, (2) their appropriate locations to locate them. The aim is to maximize performance while minimizing the delay between switches and some controllers regardless of the network traffic variations. The placement problem in a distributed controller design is referred to as the *controller placement problem (CPP)* that estimates the minimum number of controllers, and their placements in the network [396].

Concerning NFV, placing the NFVO and VNFM in a large-scale distributed system is a very challenging task, due to the negative impact on performance and operational cost [372]. Therefore, efficient placement of VNF nodes on the NFVI especially for scenarios such as users served by mobile IoT-enabled devices or users traveling in train at high speed plays a crucial role in reducing latency and further improving the end-users' QoE. The goal of VNF placement problem [397] also known as the VNF embedding [398] is to find the optimal location for either a single type of VNF or a set of various VNFs while performance levels (e.g., network load, resource allocation, and power consumption) for ensuring the end users' QoS/QoE is guaranteed. Taking into account the fact that future networks are anticipated to be supported with softwarized infrastructure deployment, the placement of the MANO functional blocks (e.g., VNFMs and VNFs) described in section IV-B1 is indeed an important area that needs significant research efforts.

Although some of the current placement and optimization solutions consider techniques such as open search and performance metrics such as mobility, network load, QoS/QoE or latency as shown in Table X, these proposals still do not take into account the joint placement problems of SDN controller and that of VNFs and VNFMs. Open search allows the SDN controller to be placed in any location within the geographical area of switches in the network. Given the location of the switches in the open search technique, the entire region of switches is searched to find the optimal placement of controllers. We believe that it would be of interest if these two placement problems can be jointly considered together by developing a QoE-aware multi-objective optimization model considering both SDN controllers and VNFs/VNFMs placement in future networks. This is so because placement planning of the controller or the VNFs separately could still affect the performance and reliability of the SDN/NFV-based system, which in turn can negatively degrade the end-users' QoE.

### G. Multimedia Communications in Internet of Things (IoTs)

Inevitably, the imminent arrival of the IoT, consisting of interconnected devices, creates scalability, mobility, security and privacy, QoE resource management [399] and multimedia network management challenges [400], [401]. Authors in [402] present QoE management aspects of multimedia in IoT applications and define a layered QoE model aimed at evaluating and estimating the overall QoE. Research challenges associated with big multimedia data, such as scalability, accessibility, reliability, heterogeneity, and QoS/QoE requirements are addressed by Kumari *et al.* in [403]. Low-complexity encoding, resilience to transmission errors, high data rate with low-power and delay bound are some of the stringent requirements on video codecs for IoT as described in [404]. Long *et al.* [405] provide an edge computing framework to enable cooperative processing on mobile devices for delay-sensitive multimedia IoT tasks. A new architecture based on the concept of Quality of Things (QoT) for multimedia communications in IoTs is proposed in [406], together with its challenges and future research directions. Some of the issues related to video traffic prioritization, virtualization of network elements, mobility management, and security are highlighted in [407]. With the current trend of softwarization and virtualization towards 5G networks, it would be essential to investigate how QoE of IoT can be managed in SDN and NFV.

TABLE X: The Key research works related to scalability, resilience, and optimization in SDN and NFV.

| Technique | Large Scale Network | Open Search | Performance Metrics | | |
|---|---|---|---|---|---|
| | | | Latency | Network Load | QoS/QoE |
| Huque et al. [408] | ✓ | ✓ | ✓ | Dynamic | ✗ |
| Bari et al. [409] | ✓ | ✗ | ✓ | Dynamic | ✗ |
| Jimenez et al. [80] | ✓ | ✗ | ✓ | Static | ✓ |
| Yao et al. [410] | ✓ | ✗ | ✓ | Static | ✗ |
| Lange et al. [411] | ✓ | ✗ | ✓ | ✗ | ✓ |
| Rath et al. [412] | ✗ | ✗ | ✓ | Static | ✗ |
| Sallahi et al. [413] | ✗ | ✗ | ✓ | Static | ✗ |
| Heller et al. [396] | ✗ | ✗ | ✓ | ✗ | ✗ |
| Hu et al. [395] | ✗ | ✗ | ✓ | ✗ | ✓ |
| Hock et al. [414] | ✗ | ✗ | ✓ | ✗ | ✓ |

*H. Summary*

We summarize in Table VIII the QoE management challenges for multimedia services in future softwarized networks. We also identify some research opportunities for future explorations, especially in the area of QoE management and orchestration of resources, HTTP adaptive streaming using QUIC/MPTCP, transport layer protocols, emerging multimedia services and applications (e.g., AR/VR, 4K/8K videos) and OTTP-ISP collaborative service management. We note that despite the recent efforts towards overcoming QoE-control and management challenges today, and the rapid evolution of SDN and NFV towards future communication systems, there are still significant gaps in some areas that need extensive research and investigations as summarized in Table IX. In order to cope with the speed at which SDN and NFV are being proposed by service providers and MNO for QoE-provisioning to the end-users, more research pertaining to multimedia services should be conducted in the following vital areas that have not been explored extensively in the past: QoE-driven network resource sharing and slicing, QoE business models in softwarized infrastructures, intelligent QoE-based big data strategies, scalability, resilience, and optimization in SDN/NFV, network performance, evaluation and benchmarks, new security, privacy and trust QoE-based models in future softwarized 5G networks.

IX. CONCLUSION

The exponential growth of Internet traffic due to the rising popularity of the multimedia services over the Internet has created network resources management concern for ISPs and OTTP. This is due to an inefficient utilization of available network resources, and the huge pressure on both the ISPs and OTTP to provide service with good quality to the end users. The academia and the industry are embracing SDN and NFV as future technologies to help overcome these challenges. SDN and NFV promise to provide and implement new capabilities and solutions for enabling future networks (e.g., 5G) control to be adaptable, programmable and cost-effective. However, due to limited network resources, it is a challenge for the service providers to provide high-quality multimedia services to all their customers.

Towards this end, we presented in this paper a comprehensive survey of QoE management solutions using SDN and NFV in current and future 5G networks. We started with a tutorial on the background of QoE modeling and assessment followed by a discussion of QoE monitoring, measurement, optimization and control. To introduce the reader to the latest and widely use streaming technology, a description of multimedia streaming services over the Internet with an emphasis on the HAS based applications and the recent SAND standard was provided.

We also presented the state-of-the-art of past and ongoing works in the field of SDN and NFV regarding their design considerations and implementations. We further highlighted the ongoing research projects, standardization activities and use cases related to SDN and NFV. Next, we presented a comprehensive survey of QoE management of multimedia streaming services based on different classifications such as server and network-assisted optimization approaches, QoE-centric routing in SDN/NFV, etc. The survey also extensively explored QoE-aware/driven adaptive streaming solutions using emerging architectures using MEC, Fog/Cloud and ICN. We also extend the QoE management approaches to newer domains such as immersive augmented and virtual reality, mulsemedia and video gaming applications.

Based on the survey on the QoE management aspect of multimedia services, we presented and discussed future needed research activities in the following directions: QoE-oriented network sharing and slicing, QoE business models in softwarized infrastructures, intelligent QoE-based big data strategies, scalability, resilience, and optimization in SDN/NFV and new security, privacy and QoE-based trust models in future softwarized 5G networks. We believe that the tutorial and survey provided in this paper, along with the outlined research gaps will help the readers to obtain an overview of the current work and needed future works.

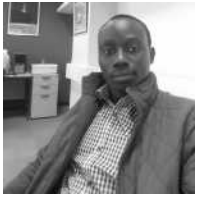

**Alcardo Alex Barakabitze** received his PhD in computing and communications from Plymouth University, UK in 2019. He received the degree in Computer Science with Honours from the University of Dar es Salaam, Tanzania in 2010 and Master Degree of Electronics and Communication Engineering with first class from Chongqing University, PR China, in May 2015. He worked as Marie Curie Fellow (2015-2018) in MSCA ITN QoE-Net. He was recognised as outstanding International Graduate Student of Chongqing University, China in 2015 due to his excellent performance. He has numerous publication in International peer-reviewed conferences and journals. Barakabitze has served as a session chair of the Future Internet and Next Generation Network (NGN) Architectures during the IEEE International Conference on Communications in Kansas City, USA, 2018. Dr. Barakabitze is a Reviewer for various journals and serves on technical program committees of leading conferences focusing on his research areas. His research interests are 5G, network management, video streaming services, SDN and NFV.

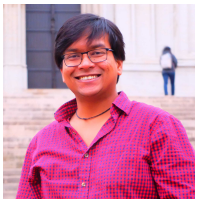

**Nabajeet Barman** (M'19) completed his Bachelor of Technology degree in Electronics Engineering in 2011 from National Institute of Technology, Surat, India and Masters in Information Technology, specializing in Communication Engineering and Media Technology from Universität Stuttgart, Germany in 2014. He worked at Bell Labs, Stuttgart, Germany, for 15 months as part of his internship and Masters thesis. He obtained his PhD in 2019 from Kingston University London as part of MSCA ITN QoE-Net and is currently a Post Doctoral Research Fellow in the Wireless Multimedia and Networking Research Group (WMN) in Kingston University. He is currently Video Quality Expert Group (VQEG) board member as part of Computer Graphics Imagery (CGI) project and is also involved in ITU-T standardization activities. His research interests include multimedia communications, machine learning and more recently distributed ledger technologies.

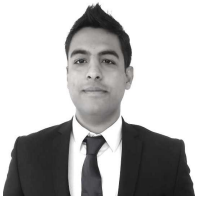

**Arslan Ahmad** (M'18) is a Ph.D. in Electronics and Computer Engineering. Currently, he is working as Senior R&D Engineer at IS-Wireless, Poland. He received his engineering degree in Aviation Electronics (Avionics Engineering) in 2011 from PAF College of Aeronautical Engineering, National University of Sciences and Technology, Pakistan. In 2014, he received Master's degree in Computational Sciences and Engineering from Research Center for Modeling & Simulation, National University of Sciences and Technology, Pakistan. He received his Ph.D. degree in Electronics and Computer Engineering from the Department of Electrical and Electronics Engineering, University of Cagliari, Italy in 2019. He has worked as Marie Curie Fellow (2015-2018) in MSCA ITN QoE-Net. He has served as a Postdoctoral researcher at University of Cagliari from 2018 to 2019. He has numerous publication in International peer-reviewed conferences and journals, two of which has been awarded Best paper award at IFIP/IEEE IM 2017 and IEEE Multimedia Technical Committee. He has been the reviewer of IEEE Transaction on Multimedia, IEEE Transaction on Network and Service Management, Springer Multimedia Tools and Application, and Elsevier Image Communication. He has been Web Chair of the 10th International Conference on Quality of Multimedia Experience (QoMEX 2018) and Technical Program Committee member of IEEE ICC and IEEE WCNC. His research interests include multimedia communication, software-defined networking, machine learning, Quality of Experience (QoE) based service and network monitoring and management. The primary focus of his research is the QoE monitoring and management of the Over-The-Top (OTTs) multimedia services and OTT-ISP collaboration for the joint multimedia service management.

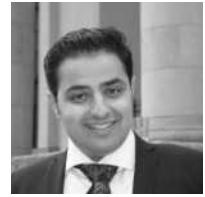

**Saman Zadtootaghaj** is a researcher at the Quality and Usability Lab at Technische Universität Berlin working on modeling the gaming quality of experience under the supervision of Prof. Dr.-Ing. Sebastian Möller. His main interest is subjective and objective quality assessment of Computer-Generated content. He received his bachelor degree from IASBS and master degree in information technology from University of Tehran. He worked as a researcher at Telekom Innovation Laboratories of Deutsche Telekom AG from 2016 to 2018 as part of European project called QoE-Net. He is currently the chair of Computer-Generated Imagery (CGI) group at Video Quality Expert Group (VQEG).

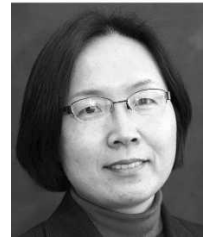

**Lingfen Sun** is an Associate Professor (Reader) of Multimedia Communications and Networks at the School of Computing, Electronics and Mathematics, Plymouth University, UK. She received the Ph.D. degree in computing and communications from Plymouth University. She has led the UoP team in several EU FP7/H2020 and industry- funded projects related to multimedia QoE including ADAMANTIUM, GERYON, Qualinet and QoENet. She is a Fellow of the Higher Education Academy (HEA) and a member of IEEE as well as an Associate Editor for IEEE Transactions on Multimedia (2016 - 2018). She was a Symposium Chair for Communication Software, Services and Multimedia Applications (CSSMA) for IEEE ICC 2014, IEEE ICME 2011 and a Post & Demo Co-Chair of IEEE Globecom, 2010. She was a member of the Award Board for IEEE MMTC (2012 - 2014) and a chair of QoE Interest Group (QoEIG), Multimedia Communications Technical Committee (MMTC) of IEEE Communication Society (2010 - 2012). She has authored or co-authored 80 peer-refereed technical papers since 2000 and filed one patent. Her current research interests include multimedia quality assessment, QoS/QoE management and control, VoIP, and network performance characterization.

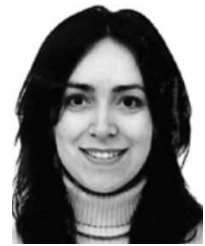

**Maria G. Martini** is a Professor in the Faculty of Science, Engineering and Computing at Kingston University, London, where she also leads the Wireless Multimedia Networking Research Group. She received the Laurea in electronic engineering (summa cum laude) from the University of Perugia (Italy) in 1998 and the Ph.D. in Electronics and Computer Science from the University of Bologna (Italy) in 2002. She has led the KU team in a number of national and international research projects, funded by the European Commission (e.g., OPTIMIX, CONCERTO, QoE-NET, Qualinet), UK research councils, UK Technology Strategy Board / InnovateUK, and international industries. An IEEE Senior Member (since 2007) and Associate Editor for IEEE Transactions on Multimedia (2014-2018), she has also been lead guest editor for the IEEE JSAC special issue on "QoE-aware wireless multimedia systems" and guest editor for the IEEE Journal of Biomedical and Health Informatics, IEEE Multimedia, and the Int. Journal on Telemedicine and Applications, among others. She chaired / organized a number of conferences and workshops. She is part of international committees and expert groups, including the NetWorld2020 European Technology Platform expert advisory group, the Video Quality Expert Group (VQEG) and the IEEE Multimedia Communications technical committee, where she has served as vice-chair (2014-2016), as chair (2012-2014) of the 3D Rendering, Processing, and Communications Interest Group, and as key member of the QoE and multimedia streaming IG. She is Expert Evaluator for the European Commission and EPSRC among others. Her research interests include QoE-driven wireless multimedia communications, decision theory, video quality assessment, and medical applications. She has authored about 150 scientific articles, contributions to standardization groups (IEEE, ITU), and several patents on wireless video.

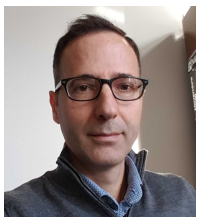

**Luigi Atzori** (SM'09) is Professor at the Department of Electrical and Electronic Engineering at the University of Cagliari (Italy), where he leads the laboratory of Multimedia and Communications with around 15 affiliates (http://mclab.diee.unica.it). L. Atzori research interests are in multimedia communications and computer networking (wireless and wireline), with emphasis on multimedia QoE, multimedia streaming, NGN service management, service management in wireless sensor networks, architecture and services in the Internet of Things. He has been the area and guest editor for many journals, including ACM/Springer Wireless Networks Journal, IEEE Communications Magazine, Springer Monet, Elsevier Signal Processing: Image Communications Journals, IEEE IoT Journal, Elsevier Ad Hoc Networks. He served as a technical program chair for various international conferences and workshops. He served as a reviewer and panelist for many funding agencies, including H2020, FP7, Cost and Italian MIUR.